\documentclass[12pt]{article}

\usepackage[dvips]{epsfig}
\usepackage{amsfonts}

\usepackage{amsmath}
\usepackage{amssymb}
\allowdisplaybreaks
\usepackage{cases}

\begin{document}

\baselineskip=16.4pt plus 0.2pt minus 0.1pt

%%%%%%%%%% Private macros %%%%%%%%%%%%

\makeatletter
\@addtoreset{equation}{section}
\renewcommand{\theequation}{\thesection.\arabic{equation}}

\newcommand{\al}{\alpha}
\newcommand{\ap}{\alpha'}
\newcommand{\sap}{\sqrt{\alpha'}}
\newcommand{\la}{\lambda}
\newcommand{\p}{\partial}
\newcommand{\be}{\begin{equation}}
\newcommand{\ee}{\end{equation}}
\newcommand{\nn}{\nonumber}
\newcommand{\ds}{\displaystyle}
\newcommand{\Pmatrix}[1]{\begin{pmatrix} #1 \end{pmatrix}}
\newcommand{\wt}[1]{\widetilde{#1}}
\newcommand{\wh}[1]{\widehat{#1}}
\newcommand{\bm}[1]{\boldsymbol{#1}}
\newcommand{\diag}{\mathop{\rm diag}}
\newcommand{\bra}[1]{\langle\, #1\,\vert}
\newcommand{\ket}[1]{\vert\, #1\,\rangle}
\newcommand{\braket}[2]{\langle\, #1\,\vert\, #2\,\rangle}
\newcommand{\Drv}[2]{\frac{\p #1}{\p #2}}
\newcommand{\abs}[1]{\left\vert #1\right\vert}
\newcommand{\cN}{\mathcal{N}}
\newcommand{\cO}{\mathcal{O}}
\newcommand{\cQ}{\mathcal{Q}}
\newcommand{\hhQ}{\wh{\wh{Q\,}}}
\newcommand{\VEV}[1]{\langle #1\rangle}
\newcommand{\tpart}{t_{\mbox{\scriptsize one-period}}}
\newcommand{\Res}{\mathop{{\rm Res}}}
\renewcommand{\Re}{\mathop{{\rm Re}}}
\newcommand{\calO}{O}
\newcommand{\Zop}{Z_{\mbox{\scriptsize 1-pair}}}
\newcommand{\CV}{C_V}
\newcommand{\DltH}{\Delta H(n)}
%%%%%%%%%% End of private macros %%%%%%%%%%%%

\begin{titlepage}
\title{
\hfill\parbox{4cm}
{\normalsize KUNS-1897\\{\tt hep-th/0403031}}\\
\vspace{1cm}
{\bf Rolling Tachyon Solution\\ in Vacuum String Field Theory
}
}
\author{
Masako {\sc Fujita}\thanks{
{\tt masako@gauge.scphys.kyoto-u.ac.jp}}
\
and
\
Hiroyuki {\sc Hata}\thanks{
{\tt hata@gauge.scphys.kyoto-u.ac.jp}}
\\[15pt]
{\it Department of Physics, Kyoto University, Kyoto 606-8502, Japan}
}
\date{\normalsize March, 2004}
\maketitle
\thispagestyle{empty}
\begin{abstract}
\normalsize

We construct a time-dependent solution in vacuum string field theory
and investigate whether the solution can be regarded as a rolling
tachyon solution.
First, compactifying one space direction on a circle of radius $R$,
we construct a space-dependent solution given as an infinite number of
$*$-products of a string field with center-of-mass
momentum dependence of the form $e^{-b p^2/4}$.
Our time-dependent solution is obtained by an inverse Wick rotation
of the compactified space direction.
We focus on one particular component field of the solution, which
takes the form of the partition function of a Coulomb system on a
circle with temperature $R^2$.
Analyzing this component field both analytically and numerically using
Monte Carlo simulation, we find that the parameter $b$ in the solution
must be set equal to zero for the solution to approach a finite value
in the large time limit $x^0\to\infty$.
We also explore the possibility that the self-dual radius $R=\sap$ is
a phase transition point of our Coulomb system.

\end{abstract}
\end{titlepage}

\section{Introduction}

The rolling tachyon process  represents the decay of unstable D-branes
in bosonic and superstring theories
\cite{rolling,Callan}.
This process is described in the limit of vanishing string coupling
constant by an exactly solvable boundary conformal field theory (BCFT).
Study of this process has recently evolved into various interesting
physics including open-closed duality at the tree level, a new
understanding of $c=1$ matrix theory and Liouville field theory,
and the rolling tachyon cosmology
(see \cite{OCduality,Klebanov,tcosmology} and the references therein).
However, there still remain many problems left unresolved; in
particular, the closed string emission and its back-reaction
\cite{OkudaSugimoto}.

One may think that such problems can be analyzed using string field
theory (SFT), which is a candidate of nonperturbative
formulation of string theory and has played critical roles in the
study of static properties of tachyon condensation
(see \cite{Ohmorireview,TZreview} and the references therein).
However, SFT has not been successfully applied to the time-dependent
rolling tachyon process.
The main reason is that no satisfactory classical solution
representing the rolling process has been known in SFT, though
there have appeared a number of approaches toward the construction of
the solutions
\cite{padic,Kluson,Yang,Aref'eva,Volovich,Fujita,Ohmori}.
Among such approaches, refs.\ \cite{padic,Fujita} examined
time-dependent solutions in cubic string field theory (CSFT)
\cite{Witten-SFT} by truncating the string field to a few lower mass
component fields and expanding them in terms of the modes $e^{n x^0}$
($n=0,\pm 1,\pm 2,\cdots$).

Let us summarize the result of our previous paper \cite{Fujita} (we
use the unit of $\ap=1$).
We expanded the tachyon component field $t(x^0)$ as
\be
t(x^0)=\sum_{n=0}^{\infty} t_n \cosh nx^0\,,
\label{tachyon}
\ee
and solved the equation of motion for the coefficients $t_n$
numerically (by treating $t_1$ as a free parameter of the solution).
Our analysis shows that the $n$-dependence of $t_n$ is given by
\be
t_n\sim\lambda^{-n^2} \left(-\beta\right)^{n} \,,
\label{CSFTresult}
\ee
up to a complicated subleading $n$-dependence. Here, $\lambda$ is a
constant $3^{9/2}/2^6$ and $\beta$ is a parameter related to $t_1$.
{}From the effective field theory analysis, the rolling tachyon solution
is expected to approach the stable non-perturbative vacuum at large
time $x^0\to\infty$ \cite{tmatter}.
If $t_n$ behaves like \eqref{CSFTresult}, however, the profile of the
tachyon field $t(x^0)$ cannot be such a desirable one: it oscillates
with rapidly growing amplitude (see also \eqref{tn0behavior} and
\eqref{evalt(x)}):
\be
t(x^0) \sim e^{(x^0)^2/(4\ln\lambda)}\times
({\mbox{oscillating term}})\,.
\ee
Since the radius of convergence with respect to $x^0$ of the series
\eqref{tachyon} is infinite for $t_n$ of \eqref{CSFTresult},
we cannot expect that analytic continuation gives another $t(x^0)$
which converges to a constant as $x^0\to\infty$.

In order for the series \eqref{tachyon} to reproduce a desirable
profile, it is absolutely necessary that the fast dumping
factor $\lambda^{-n^2}$ of \eqref{CSFTresult} disappears.
If this were the case and, in addition, if $t_n$ were exactly given by
\be
t_n=(-\beta)^n\, ,
\label{tn=beta^n}
\ee
analytic continuation of the series \eqref{tachyon} would lead to
\be
t(x^0)=-1+\frac{1}{1+\beta e^{x^0}}+\frac{1}{1+\beta e^{-x^0}}
\,,
\label{analyticcont}
\ee
which approaches monotonically a constant as $x^0\to\infty$.
This particular $t(x^0)$ has another desirable feature that it becomes
independent of $x^0$ when $\beta=0$ and $1$, which may correspond
to sitting on the unstable vacuum and the stable one, respectively.
Since CSFT should reproduce the rolling tachyon process, it is
expected that the behavior \eqref{CSFTresult} is an artifact of
truncating the string field to lower mass component fields and that
some kind of more sensible analysis would effectively realize
$\lambda=1$.

The purpose of this paper is to study the rolling tachyon solution in
vacuum string field theory (VSFT) \cite{RSZ1,RSZ2,RSZ3,GRSZ},
which has been proposed as a candidate SFT expanded around the stable
tachyon vacuum.
The action of VSFT is simply given by that of CSFT with the BRST
operator $Q_{\rm B}$ in the kinetic term replaced by another operator
$\cQ$ consisting only of ghost oscillators. Owing to the purely ghost
nature of $\cQ$, the classical equation of motion of VSFT is
factorized into the matter part and the ghost one, each of which can
be solved analytically to give static solutions representing
D$p$-branes. In fact, analysis of the fluctuation modes around the
solution has successfully reproduced the open string spectrum at the
unstable vacuum although there still remain problems concerning the
energy density of the solution
\cite{HataVSFT,RSZnote,Okawa}.\footnote{
See also \cite{Okawasome,Bonora} for recent attempts to this problem.
}
If we can similarly construct time-dependent solutions in VSFT
without truncation of the string field, we could study more reliably
whether SFT can reproduce the rolling tachyon processes,
and furthermore, the unresolved problems mentioned at the beginning of
this section.

Our strategy of constructing a time-dependent solution in VSFT is as
follows. First we prepare a lump solution depending on one space
direction which is compactified on a circle of radius $R$. Then, we
inverse-Wick-rotate this space direction to obtain a time-dependent
solution following the BCFT approach \cite{rolling}.
The lump solution of VSFT localized in uncompactified
directions has been constructed in the oscillator formalism by
introducing the creation/annihilation operators for the zero-mode in
this direction \cite{RSZ2}.
In the compactified case, however, we cannot directly apply this
method. We instead construct the matter part $\Phi^{\rm m}$ of a lump
solution as an infinite number of $*$-products of a string field
$\Omega_b$; $\Phi^{\rm m}=\Omega_b *\Omega_b *\cdots *\Omega_b$
(the ghost part is the same as that in the static solutions).
Since the equation of motion of $\Phi^{\rm m}$ is simply
$\Phi^{\rm m}*\Phi^{\rm m}=\Phi^{\rm m}$, this gives a solution if the
limit of an infinite number of $*$-product exists \cite{RZ}.
As the constituent $\Omega_b$, we adopt the one which is the
oscillator vacuum with respect to the non-zero modes and has the
Gaussian dependence $e^{-b\,p^2/4}$ on the zero-mode momentum $p$ in
the compactified direction. Finally, our time-dependent solution is
obtained by making the inverse Wick rotation of the compactified
direction $X\to -iX^0$ or $-iX^0 +\pi R$.

After constructing a time-dependent solution in VSFT, our next task is
to examine whether it represents the rolling tachyon process.
Our solution consists of an infinite number of string states, and we
focus on one particular component field $t(x^0)$ (we adopt the same
symbol as the tachyon field in the CSFT analysis).
This $t(x^0)$ has the expansion \eqref{tachyon} with $\cosh nx^0$
replaced by $\cosh(nx^0/R)$.
What is interesting about $t(x^0)$ is that it takes the form of the
partition function of a statistical system of charges at sites
distributed with an equal spacing on a unit circle. The temperature of
this system is $R^2$.
The charges interact through Coulomb potential and they also have a
self energy depending on the parameter $b$ of $\Omega_b$.
The partition function is obtained by summing over the integer value
of the charge on each site keeping the condition that the total charge
be equal to zero.

What we would like to know about $t(x^0)$ are particularly the
following two:
\vspace{-2mm}
\begin{itemize}
\item Whether $t(x^0)$ has a profile which converges to a constant as
  $x^0\to\infty$.

\item Whether the critical radius $R=1$ in the BCFT approach
 \cite{marginal,critical} is required also in our solution.
\end{itemize}
\vspace{-2mm}
That we have to put $R=1$ in our solution is also natural in view of
the fact that the correct value $-1$ of the tachyon mass squared is
reproduced from the fluctuation analysis around the D25-brane solution
of VSFT \cite{HataVSFT,RSZnote,Okawa}. For these two problems,
we carry out analysis using both analytic and numerical methods.
In particular, we can apply the Monte Carlo simulation since $t(x^0)$
is the partition function of a Coulomb system on a circle.
We find that the coefficient $t_n$ of our VSFT solution has a similar
$n$-dependence to \eqref{tachyon} with $\lambda$ depending on the
parameter $b$. This implies that the profile of $t(x^0)$ is again an
unwelcome one for a generic value of $b$: it is an oscillating
function of $x^0$ with growing amplitude.
However, we can realize $\lambda=1$ by putting $b=0$ and taking
the number of $\Omega_b$ in $\Phi^{\rm m}=\Omega_b *\cdots *\Omega_b$
to infinity by keeping this number even.
These properties seems to hold for any value of $R$.
In order to see whether $R=1$ has a special meaning for our solution,
we study the various thermodynamic properties of the Coulomb system
$t(x^0)$. First we argue using a naive free energy analysis that
there could be a phase transition at temperature
$R^2=1$. Below $R^2=1$ only the excitations of neutral boundstates of
charges are allowed, but above $R^2=1$ excitations of isolated charges
dominate the partition function.
We carry out Monte Carlo study of the internal energy and the specific
heat of the system, but cannot confirm the existence of this phase
transition. However, we find that the correlation function of the
charges show qualitatively different behaviors between the large and
small $R^2$ regions when $b=0$, possibly supporting the existence of
the phase transition.

The rest of this paper is organized as follows.
In section 2, first briefly reviewing VSFT and its classical solutions
representing various D$p$-branes, we construct time-dependent
solutions following the strategy mentioned above.
In section 3, we investigate the profile of the component field
$t(x^0)$ both analytically and numerically.
In section 4, we argue that our solution with $b=0$ could give a
rolling tachyon solution.
In section 5, we study a possible phase transition at $R^2=1$ through
various thermodynamic properties of the system.
The final section (section 6) is devoted to a summary and discussions.
In the appendix we present a proof concerning the minimum energy
configuration of the Coulomb system.

\section{Construction of a time-dependent solution in VSFT}

In this section, we shall construct a time-dependent solution
in VSFT.
As stated in section 1, we first construct a lump
solution which is localized in one spatial direction compactified
on a circle of radius $R$.
This solution is given as an infinite number of $\ast$-products of a
string field $\Omega_b$; $\Omega_b *\Omega_b *\cdots *\Omega_b$.
Our time-dependent solution is obtained by inverse-Wick-rotating
the spatial direction to the time one.
Throughout this paper, we use the convention $\ap =1$.

\subsection{D$p$-brane solutions in VSFT}

In this subsection, we briefly review the construction of lump
solutions in VSFT describing various D$p$-branes in the uncompactified
space \cite{RSZ2}.
VSFT is a string field theory around the non-perturbative vacuum where
there are only closed string states. Its action is written as
follows using the open string field $\Psi$:
\begin{align}
S&=-\frac12 \Psi\cdot \cQ\Psi -\frac13 \Psi\cdot(\Psi\ast\Psi)\nn\\
&=-\frac12 \,\bra{\Psi}\,\cQ\,\ket{\Psi}
-\frac13
{}_0\bra{\Psi}{}_1\bra{\Psi}{}_2\braket{\Psi}{V_3}_{012}.
\label{VSFTaction}
\end{align}
The BRST operator $\cQ$ of VSFT consists of only ghost operators, and
it has no non-trivial cohomology.
The three-string vertex $\ket{V_3}$ represents the mid-point
interaction of three strings, and it factorizes into the direct
product of the matter part and the ghost one.
More generically, the matter part of the $N$-string vertex $\ket{V_N}$
representing the symmetric mid-point interaction of $N$-strings
($N=3,4,\cdots$) is given by \cite{GJ,LPP}
\begin{align}
&\ket{V_N^{\rm m}}_{01\cdots N-1}
=\int\! d^{26}p_0 \cdots \!\int\! d^{26}p_{N-1}
\,\delta^{26}\!\biggl(\sum_{r=0}^{N-1} p_r\biggr)
\nn\\
&\ \times
\exp\!\left(\!-\eta_{\mu\nu}\sum_{r,s=0}^{N-1}\left[
\frac12\sum_{n,m=1}^{\infty}
V_{nm}^{rs} a_n^{(r)\mu\dag} a_m^{(s)\nu\dag}
+\sum_{n=1}^{\infty}V_{n0}^{rs} a_n^{(r)\mu\dag}p_s^\nu
+\frac12 V_{00}^{rs} p_r^\mu p_s^\nu\right]
\right)\bigotimes_{r=0}^{N-1}\ket{0;p_r}_r
\,,
\label{vertex}
\end{align}
where $\ket{0;p_r}$ is Fock vacuum of the $r$-th string carrying the
center-of-mass momentum $p_r$ (the index $r$ specifying the $N$
strings runs from $0$ to $N-1$).
Here we use the same convention as \cite{RSZ1,RSZ2,RSZ3,GRSZ}.
$a_n^{\mu(r)}$ are the matter oscillators of non-zero modes normalized
so that their commutation relations are
\be
[a_n^{(r)\mu},a_m^{(s)\nu\dag}]=\eta^{\mu\nu}\delta_{nm}\delta^{rs}\,,
\quad (n,m\geq 1)\,.
\label{CR}
\ee
The coefficients $V_{nm}^{rs}$ are called the Neumann coefficients.
In particular, $V_{00}^{rs}$ is given by
\be
V_{00}^{rs}=
\begin{cases}
\ds -\ln\abs{2 \sin\frac{\pi(r-s)}{N}}\, , & (r\ne s)\, ,
\\[4mm]
\ds 2\ln\!\left(\frac{N}{4}\right)\, , & (r=s)\, .
\end{cases}
\ee
Note that $V_{nm}^{rs}$ depends on $N$ although we do not write it
explicitly.

The action \eqref{VSFTaction} leads to the equation of motion
\be
\cQ\Psi=-\Psi\ast\Psi\,.
\ee
Assuming that the solution is given as a product of the matter part
and the ghost one, $\Psi=\Psi^{\rm m}\otimes\Psi^{\rm g}$,
the equation of motion is reduced to
\begin{align}
\Psi^{\rm m} &= \Psi^{\rm m}\ast^{\rm m}\Psi^{\rm m}\, ,
\label{mequation}\\
\cQ\Psi^{\rm g} &= -\Psi^{\rm g}\ast^{\rm g}\Psi^{\rm g} \,,
\end{align}
where $\ast^{\rm m}$ ($\ast^{\rm g}$) is the $*$-product in the matter
(ghost) sector.
In this paper, we assume that the ghost part $\Psi^{\rm g}$ is
common to the various solutions, and focus on the matter part equation
\eqref{mequation}.

Classical solutions of \eqref{mequation} which represent the
various D$p$-branes in spacetime are given in \cite{RSZ2}.
Let us review the two ways of constructing classical solutions
representing the translationally invariant D25-brane.
One way is to assume that $\Psi^{\rm m}$ is given in the form of
a squeezed state, the exponential of an oscillator bilinear acting on
the vacuum:
\be
\ket{\Psi^{\rm m}}={\cN}\exp
\left( -\frac12\eta_{\mu\nu}\sum_{m,n=1}^\infty S_{mn}a_m^{\mu\dag}
a_n^{\nu\dag}\right)\ket{0,0}\,,
\label{sol}
\ee
where $\cN$ is a normalization factor.
The equation of motion \eqref{mequation} is reduced to an algebraic
equation for the infinite dimensional matrix $S_{mn}$,
which, under a certain commutativity assumption and using the
algebraic relations among the Neumann coefficients $V_{mn}^{rs}$
\cite{GJ}, can be solved to give $S_{mn}$ in terms of $V_{mn}^{rs}$
\cite{Kostelecky}:
\be
S = CT, \qquad T=\frac{1}{2X}\left(
1 + X - \sqrt{(1+3 X)(1 - X)}\right)\, ,
\label{sol_S}
\ee
with the matrices $C$ and $X$ given by
\be
C_{mn} = (-1)^m \delta_{mn},
\qquad
X = C V^{11}\, .
\ee

Another way is to construct $\Psi^{\rm m}$ as the sliver state \cite{RZ}.
Defining the wedge states as
\be
\ket{N}_0=\underbrace{
\ket{0;0}\ast\ket{0;0}\ast\cdots\ast\ket{0;0}}_{N-1}
={}_1\bra{0;0}{}_2\bra{0;0}\cdots{}_{N-1}
\braket{0;0}{V_N}_{01\cdots N-1}\,,
\ee
they satisfy the following property:
\be
\ket{N}\ast\ket{M}=\ket{N+M-1}\,.
\label{prop_wedge}
\ee
Taking the limit $N,M\to\infty$, we have
\be
\ket{\infty}\ast\ket{\infty}=\ket{\infty}\,.
\label{prop_sliver}
\ee
Namely, the state $\ket{\infty}$ (sliver state) is a solution to
\eqref{mequation}.
It has been proved that the two solutions, \eqref{sol} and
$\ket{\infty}$, are identical with each other \cite{Okuda}.

Lump solutions localized in spatial directions can be
constructed in the same way as the D25-brane solution.
Let us denote the directions transverse to the brane by $x^\al$.
In the squeezed state construction \cite{RSZ2}, we introduce the
annihilation and the creation operators for the zero-mode in the
transverse directions
by
\be
a_0^{\al}=\frac{\sqrt{b}}{2}\,\hat{p}^\al
-\frac{i}{\sqrt{b}}\,\hat{x}^\al \, ,\quad
a_0^{\al\dag}=\frac{\sqrt{b}}{2}\,\hat{p}^\al
+\frac{i}{\sqrt{b}}\,\hat{x}^\al\,,
\label{a0}
\ee
where $b$ is an arbitrary positive constant.
Since the zero-modes $a_0^{\al}$ satisfy the same commutation
relation \eqref{CR} as the non-zero modes, we define the new Fock
vacuum $\ket{\Omega_b}$ by
\be
a_n^{(r)\al}\ket{\Omega_b}=0\, ,\quad (n\geq 0)\,.
\ee
The new vacuum $\ket{\Omega_b}$ with the normalization
$\braket{\Omega_b}{\Omega_b}=1$ is expressed in terms of the momentum
eigenstates as
\be
\ket{\Omega_b}=\prod_\al \left(\frac{b}{2\pi}\right)^{1/4}
\int_{-\infty}^\infty\!dp^\al\, e^{-(b/4)(p^\al)^2}\ket{0;p_\al}\,.
\label{omega}
\ee
With these oscillators and the coordinate-dependent vacuum
$\ket{\Omega_b}$, the transverse part of the three-string vertex
$\ket{V_3}$ can be written as
\begin{align}
\exp\left(-\frac12\sum_{r,s=0,1,2}
\sum_{m,n=0}^\infty
a^{(r)\alpha\dagger}_m V^{\prime rs}_{mn}
a_n^{(s)\alpha\dagger} \right)
\ket{\Omega_b}_{012} \, ,
\end{align}
in terms of the new coefficients $V_{nm}^{\prime\, rs}$,
which satisfy the same algebraic relations as $V_{nm}^{rs}$.
Therefore, we can construct the D$p$-brane solutions just in the same
way as the D25-brane solution:
\begin{align}
\ket{\Psi^{\rm m}_p}&=\ket{\Psi^{\rm m}_{\parallel}}\otimes
\ket{\Psi^{\rm m}_{\perp}}\nn\\
&=\exp\biggr(-\frac12\,\eta_{\mu\nu}\!\sum_{m,n=1}^{\infty} S_{mn}
a^{\mu\dag}_m a^{\nu\dag}_n\biggr)\ket{0;p}
\otimes
\exp\biggr(-\frac12\sum_{m,n=0}^{\infty} S^{\,\prime}_{mn}
a^{\al\dag}_m a^{\al\dag}_n\biggr)\ket{\Omega_b}\,,
\label{sol_lump}
\end{align}
where the indices $\mu$, $\nu$ run the directions tangental to the
branes ($\mu,\nu=0,1,\cdots,25-p$), and
$S^\prime_{mn}$ is given by \eqref{sol_S} with $V^{11}$ replaced by
$V'^{11}$.
This lump solution contains one arbitrary parameter $b$, the physical
meaning of which is not known.
It has been shown that the ratio of the tensions of D$p$-brane
solutions is independent of $b$ \cite{Okuyamab}.
Later we will argue that we must choose $b=0$ to
obtain a time-dependent solution with the desirable rolling profile.

Finally, note that,
since the modified Neumann coefficients $V^{\prime\, rs}_{mn}$ satisfy
the same algebra as the original $V^{rs}_{mn}$, the transverse part
$\ket{\Psi^{\rm m}_\perp}$ of
the lump solution \eqref{sol_lump} can be written as a sliver state:
\be
\ket{\Psi^{\rm m}_\perp}=
\lim_{N\rightarrow\infty}{}_1\bra{\Omega_b}{}_2\bra{\Omega_b}\cdots
{}_{N-1}\braket{\Omega_b}{V_{N\,\perp}}_{01\cdots N-1}\, ,
\ee
where $\ket{V_{N\,\perp}}$ is the transverse part of the $N$-string
vertex.

\subsection{Time-dependent solution in VSFT}
\label{TDSinVSFT}

Now let us construct a time-dependent solution in VSFT which possibly
represents the process of rolling tachyon.
This consists of the following two steps:
\begin{itemize}
\item Construction of a lump solution of VSFT localized in one space
  direction which is compactified on a circle of radius $R$.

\item Inverse Wick rotation of the compactified space
  direction to the time one on this lump solution to obtain a
  time-dependent solution in VSFT.
\end{itemize}
Since both the solution and the string vertices have
factorized forms with respect to the spacetime directions, we shall
focus only on this transverse direction of the brane in the rest of
this paper.

First, we shall construct a lump solution on a circle. The squeezed
state construction explained in the previous
subsection, however, cannot be directly applied to the compactified
case since the zero-mode creation/annihilation operators of \eqref{a0}
are ill-defined due to the periodicity
$\hat{x}^\al\sim \hat{x}^\al+2\pi R$.
Therefore, we shall adopt the sliver state construction of the lump
solution. Namely, let us consider
\be
\lim_{N\to\infty}
\underbrace{\ket{\Omega_b}*\cdots *\ket{\Omega_b}}_{N-1}
=
\lim_{N\rightarrow\infty}{}_1\bra{\Omega_b}\cdots
{}_{N-1}\braket{\Omega_b}{V_N}_{01\cdots N-1}\, ,
\label{compsliver}
\ee
with a suitably chosen $\ket{\Omega_b}$. If the limit $N\to\infty$
of \eqref{compsliver} exits, it gives a solution of VSFT.
Taking into account that the momentum zero-mode $p$ in the
compactified direction takes discrete values $p=n/R$,
we adopt as the state $\ket{\Omega_b}$ in \eqref{compsliver}
the following one which is a natural compactified version of
\eqref{omega}:
\be
\ket{\Omega_b}=\sum_{n=-\infty}^{\infty}e^{-(b/4)(n/R)^2}
\ket{0;n/R}\,,
\ee
where $\ket{0;n/R}$ is the momentum eigenstate (and the Fock vacuum of
the non-zero modes) with the normalization
$\braket{0;m/R}{0;n/R}=\delta_{n,m}$.
The $N$-string vertex $\ket{V_N^{\rm m}}$ for
the compactified direction is given by \eqref{vertex} with the
replacements:
\be
\int\! dp \rightarrow \frac{1}{R}\sum_n\,,\quad
\delta(p)\rightarrow R\,\delta_{n,0}\,,\quad
\ket{0;p}\rightarrow \sqrt{R}\,\ket{0;n/R}\, .
\ee
Then the state $\ket{\Omega_b}*\cdots *\ket{\Omega_b}$ in the
$x$-representation for the center-of-mass dependence is given by
\begin{align}
&
\bra{x}\Bigl(\underbrace{\ket{\Omega_b}*\cdots *\ket{\Omega_b}}_{N-1}
\Bigr)=
\sum_{n_0=-\infty}^{\infty}
\sum_{n_1=-\infty}^{\infty} \cdots
\sum_{n_{N-1}=-\infty}^{\infty}
\delta_{\sum_{r=0}^{N-1} n_r,0}
\exp\!\left(\frac{in_0 x}{R}\right)
\nn\\
&\times
 \exp\biggr(\!-\frac12 \sum_{n,m=1}^{\infty}
V_{nm}^{00} a_n^{\dag} a_m^{\dag}
-\sum_{s=0}^{N-1}\sum_{n=1}^{\infty}
 V_{n0}^{0s} a_n^{\dag}  \frac{n_s}{R}
-\frac{1}{2R^2} \sum_{r,s=0}^{N-1}
V_{00}^{rs} n_r n_s -\frac{b}{4R^2}\sum_{r=1}^{N-1}n_r^2
\biggr)\ket{0}\, ,
\end{align}
which in the limit $N\to\infty$ should give a lump solution on a
circle.
In this paper we are interested only in the time-dependence of the
solution and hence ignore the overall constant factor multiplying the
solution.

Our construction of a time-dependent solution of VSFT is completed by
making the inverse Wick rotation $X\rightarrow -iX^0$, namely,
$x\to -ix^0$ and $a_n^\dagger\to -i a_n^\dagger$, on this lump
solution:
\begin{align}
& \ket{\Psi(x^0)}=\lim_{N\to\infty}
\sum_{n_0=-\infty}^{\infty} \sum_{n_1=-\infty}^{\infty} \cdots
\sum_{n_{N-1}=-\infty}^{\infty}
\delta_{\sum_{r=0}^{N-1} n_r,0}
\exp\!\left(\frac{n_0 x^0}{R}\right)
\nn\\
&\times
 \exp\biggr(\,\frac12 \sum_{n,m=1}^{\infty}
V_{nm}^{00} a_n^{\dag} a_m^{\dag}
+i\sum_{s=0}^{N-1}\sum_{n=1}^{\infty}
 V_{n0}^{0s} a_n^{\dag}  \frac{n_s}{R}
-\frac{1}{2R^2} \sum_{r,s=0}^{N-1}Q_{rs}n_r n_s
\biggr)\ket{0}\,,
\label{Psi}
\end{align}
where $Q_{rs}$ is defined by
\be
Q_{rs}= V_{00}^{rs}+\frac{b}{2}\,\delta_{r,s}\left(\delta_{r,0}-1\right)=
\begin{cases}
\ds -2\ln\!\abs{ 2\sin\frac{\pi (r-s)}{N}}\, ,
& (r\neq s)\, ,\\[12pt]
\ds 2\ln\!\left(\frac{N}{4}\right)+\frac{b}{2}\, ,
& (r=s\neq 0)\, ,\\[10pt]
\ds 2\ln\!\left(\frac{N}{4}\right)\, ,& (r=s=0)\, .
\end{cases}
\label{Qrs}
\ee

Taylor expansion of
$\exp\left(i\sum_s\sum_n V_{n0}^{0s}a_n^\dagger n_s/R\right)$
in \eqref{Psi} gives an expression of $\ket{\Psi(x^0)}$
as an infinite summation, $\ket{\Psi(x^0)}
=\sum_{\alpha}\ket{\alpha}\,\varphi_\alpha(x^0)$,
where $\ket{\alpha}$ are the static string states of the form
$a^\dagger\cdots a^\dagger
\exp(\frac12\sum V_{mn}^{00}a^\dag_m a^\dag_n)\ket{0}$,
and $\varphi_\alpha(x^0)$ are the corresponding time-dependent
component fields.
In this paper, we shall, for simplicity, focus on the component field
of the pure squeezed state
$\exp(\frac12\sum V_{mn}^{00}a^\dag_m a^\dag_n)\ket{0}$.
Since we have $\lim_{N\to\infty} V_{mn}^{00}=S_{mn}$ \cite{Okuda},
this component field is that for the state representing the unstable
vacuum.
Denoting this component field by $t(x^0)$, we have
\be
t(x^0)=\sum_{n_0=-\infty}^{\infty} e^{n_0x^0/R}\,t_{n_0}\,,
\label{tx}
\ee
where $t_{n_0}$ is given by
\be
t_{n_0}=\sum_{
\begin{smallmatrix}
n_1,\cdots,n_{N-1}=-\infty\\
(n_0+n_1+\cdots +n_{N-1}=0)
\end{smallmatrix}}^\infty
\exp\!\left(-\frac{1}{R^2} H(n_r;n_0)\right)\,,
\label{tn}
\ee
with
\be
H(n_r;n_0)=\frac12\sum_{r,s=0}^{N-1}Q_{rs}n_rn_s\,.
\label{H}
\ee
Note that the coefficient $t_{n_0}$ can be regarded as the partition
function of a statistical system with Hamiltonian $H$ and the
temperature $R^2$.
\begin{figure}[htbp]
\begin{center}
\epsfig{file=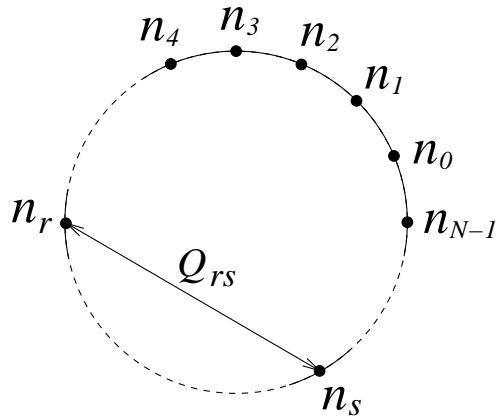, width=6.5cm}
\end{center}
\vspace{-3mm}
\caption{There are $N$ charges $n_r$ on the unit circle.
Each charge $n_r$ takes integer values, and the total charge must be
equal to zero. The charges $n_r$ and $n_s$ interact via the Coulomb
potential $Q_{rs}$.
}
\label{system}
\end{figure}
In this statistical system, we have $N$ charges $n_r$ on a unit circle
at an equal interval (figure \ref{system}). The charges $n_r$ take
integer values from
$-\infty$ to $+\infty$, and they have the self interaction $Q_{rr}$
and the two-dimensional Coulomb interaction $Q_{rs}$ ($r\ne s$)
between each other.
In $t_{n_0}$, the charge $n_0$ at $r=0$ is fixed, and there is a
constraint that the total charge be equal to zero.
Note that $t_{n_0}$ is positive definite, $t_{n_0}>0$, and is even
under $n_0\to -n_0$:
\be
t_{-n_0}=t_{n_0}\, .
\label{tniseven}
\ee
On the other hand, $t(x^0)$ itself is interpreted as the partition
function of the statistical system in the presence of the external
source $x^0/R$ for $n_0$.

Solving the constraint $\sum_{r=0}^{N-1}n_r=0$ to eliminate $n_{N-1}$,
$t_{n_0}$ and $t(x^0)$ are rewritten using independent variables
without constraint:
\be
t_{n_0}=\sum_{n_1,\cdots,n_{N-2}=-\infty}^\infty
\exp\!\left(
-\,\frac{1}{2R^2}\sum_{r,s=0}^{N-2}\wh{Q}_{rs}n_rn_s\right)\,,
\label{t_n0}
\ee
where $\wh{Q}_{rs}$ is a $(N-1)\times (N-1)$ matrix given by
\be
\wh{Q}_{r,s}=Q_{rs}-Q_{r,N-1}-Q_{N-1,s}+Q_{N-1,N-1}\, ,\quad
(r,s=0,1,\cdots,N-2)\,.
\ee
We have checked numerically that the matrix $\wh{Q}_{rs}$ is a
positive definite matrix.

In addition to the above $\ket{\Psi(x^0)}$ obtained by the inverse
Wick rotation $X\rightarrow -iX^0$,
we have another time-dependent solution via a different
inverse Wick rotation, $X\rightarrow -iX^0+\pi R$.
This new solution is obtained simply by inserting $(-1)^{n_0}$ into
\eqref{Psi} and \eqref{tx}, and satisfies the hermiticity condition.
As we shall explain in the next section, we expect that this new
solution with $(-1)^{n_0}$ represents the rolling process to the
stable tachyon vacuum, while the orignal solution given by \eqref{Psi}
represents the rolling in the direction where the potential is
unbounded from below.

\section{Analysis of the component field $\bm{t(x^0)}$}

In this section, we study the profile of the component field $t(x^0)$
given by \eqref{tx} both analytically and numerically.
If our VSFT solution \eqref{Psi} represents the rolling tachyon
solution, the component field $t(x^0)$ as well as the whole
$\ket{\Psi(x^0)}$ should approach zero, namely the tachyon vacuum, as
$x^0\rightarrow\infty$.

Let us mention the expected $n_0$-dependence of the coefficient
$t_{n_0}$, \eqref{tn} and \eqref{t_n0}, necessary for $t(x^0)$ to have
a rolling tachyon profile. Suppose that the $n_0$-dependence of
$t_{n_0}$ is given by
\be
t_{n_0}=e^{-an_0^2}\,\wt{t}_{n_0}\,,
\label{tn0behavior}
\ee
where $\wt{t}_{n_0}$ has a milder $n_0$-dependence than the leading
factor $e^{-an_0^2}$.
We expect that \linebreak
$\lim_{n\to\infty}\wt{t}_{n}/\wt{t}_{n+1}$ is finite
and larger than one, namely, that the series \eqref{tx} with $t_{n_0}$
replaced by $\wt{t}_{n_0}$ has a finite radius of convergence with
respect to $x^0$.
A typical example is $\wt{t}_{n_0}\sim e^{-b\abs{n_0}}$.
Such $t_{n_0}$ actually appeared in the time-dependent solution in
CSFT in the level-truncation approximation \cite{padic,Fujita}.
For this $t_{n_0}$, we have
\be
t(x^0)=e^{(x^0)^2/(4a)}\,\wt{t}(x^0),\qquad
\wt{t}(x^0)=\sum_{n=-\infty}^\infty
\wt{t}_n\,e^{-(x^0-2na)^2/(4a)}\,.
\label{evalt(x)}
\ee
If  $\wt{t}_{n_0}$ does not depend on $n_0$, $\wt{t}(x^0)$ is a periodic
function of $x^0$ with period $2a$, and the whole $t(x^0)$ cannot have a
desired profile: it oscillates with
blowing up amplitude $e^{(x^0)^2/(4a)}$ as $x^0\to\infty$.
Even if $\wt{t}_{n_0}$ has a mild $n_0$-dependence such as
$\wt{t}_{n_0}\sim e^{-b\abs{n_0}}$, it seems very unlikely that $t(x^0)$
approaches a finite value in the limit $x^0\to\infty$.
These properties persist in the alternating sign solution
$(-1)^{n_0}t_{n_0}$ obtained by another inverse Wick rotation
mentioned at the end of section 2.
Therefore, it is necessary that the leading term
$e^{-an_0^2}$ in \eqref{tn0behavior} is missing, namely, we must have
$a=0$.
If this is the case, the series $t(x^0)$ \eqref{tx} would have a finite
radius of convergence, and the analytic continuation would give a
globally defined $t(x^0)$ such as \eqref{analyticcont}.
Since $t_{n_0}$ is positive definite, $t(x^0)$ diverges at the
radius of convergence and corresponds to the rolling in the direction
of the unbounded potential. On the other hand, another $t(x^0)$ with
alternating sign coefficients $(-1)^{n_0}\,t_{n_0}$ is expected to be
finite at the radius of convergence and represent the rolling to the
tachyon vacuum.

In the rest of this section we shall study whether the condition $a=0$
is satisfied for the present solution.
We shall omit the indices $0$ of $n_0$ and $x^0$
unless confusion occurs.

\subsection{Analysis for $R^2\gg 1$ and $R^2\ll 1$}

In this subsection, we shall consider the $n$-dependence of the
coefficient $t_{n}$ of \eqref{t_n0} for $R^2\gg 1$ and $R^2\ll 1$.
First, for the analysis in the region $R^2\gg 1$ and also for later
use, we present another expression of $t_{n}$ obtained by applying
the Poisson's resummation formula,
\begin{equation}
\sum_{n=-\infty}^\infty g(n)=\sum_{m=-\infty}^\infty
\int_{-\infty}^\infty\! dy\,g(y)\,e^{2\pi i m y} \, ,
\end{equation}
to the $n_r$-summations in \eqref{t_n0}:
\begin{align}
t_{n}&=R^{N-2}\Bigl(\det\hhQ\Bigr)^{-1/2}
\exp\!\left(-\frac{1}{
2\bigl(\wh{Q}^{-1}\bigr)_{00}}\frac{n^2}{R^2}\right)
\nn\\
&\qquad\times
\sum_{m_1,\cdots,m_{N-2}=-\infty}^\infty
\exp\left\{
-2\pi^2 R^2\sum_{r,s=1}^{N-2} m_r\Bigl(\hhQ^{-1}\Bigr)_{rs}m_s
+ 2\pi i n\sum_{r=1}^{N-2}n_r^C m_r\right\}\,,
\label{tn0Poisson}
\end{align}
where the matrix $\hhQ$ is the lower-right $(N-2)\times(N-2)$ part of
$\wh{Q}$,
\begin{equation}
\hhQ_{rs}=\wh{Q}_{rs}\, ,\quad (r,s=1,2,\cdots, N-2)\, ,
\label{hhQ}
\end{equation}
and $n_r^C$ is defined by
\begin{equation}
n_r^C=-\sum_{s=1}^{N-2}\bigl(\hhQ^{-1}\bigr)_{rs}\wh{Q}_{s0}\,,
\quad (r=1,2,\cdots, N-2)\,.
\label{nrC}
\end{equation}
In obtaining the expression \eqref{tn0Poisson}, we have used that
\begin{equation}
\wh{Q}_{00}-\sum_{r,s=1}^{N-2}\wh{Q}_{0r}\bigl(\hhQ^{-1}\bigr)_{rs}
\wh{Q}_{s0}=\frac{1}{\bigl(\wh{Q}^{-1}\bigr)_{00}}\,,
\end{equation}
which is valid for any matrix $\wh{Q}$ and its $(N-2)\times(N-2)$
submatrix $\hhQ$.

Some comments on the formula \eqref{tn0Poisson} are in order.
First, $\{n_r^C\}$ in \eqref{nrC} is nothing but the configuration
which, without the constraint that $n_r^C$ be integers and keeping
$n=1$ fixed, minimizes the Hamiltonian \eqref{H},
\begin{equation}
H(n_r;n)=\frac12\sum_{r,s=1}^{N-2}\hhQ_{rs}n_r n_s
+n\sum_{r=1}^{N-2}\wh{Q}_{0r}n_r
+\frac12\,\wh{Q}_{00}n^2 \,.
\end{equation}
\begin{figure}[htbp]
\vspace{-0.5cm}
\begin{center}
\leavevmode
\put(203,105){{\large $\bm{r}$}}
\put(10,165){{\large $\bm{n_r^C}$}}
\put(438,149){{\large $\bm{r}$}}
\put(250,165){{\large $\bm{n_r^C}$}}
\put(80,50){{\large $\bm{b=0.1}$}}
\put(320,50){{\large $\bm{b=10}$}}
\epsfig{file=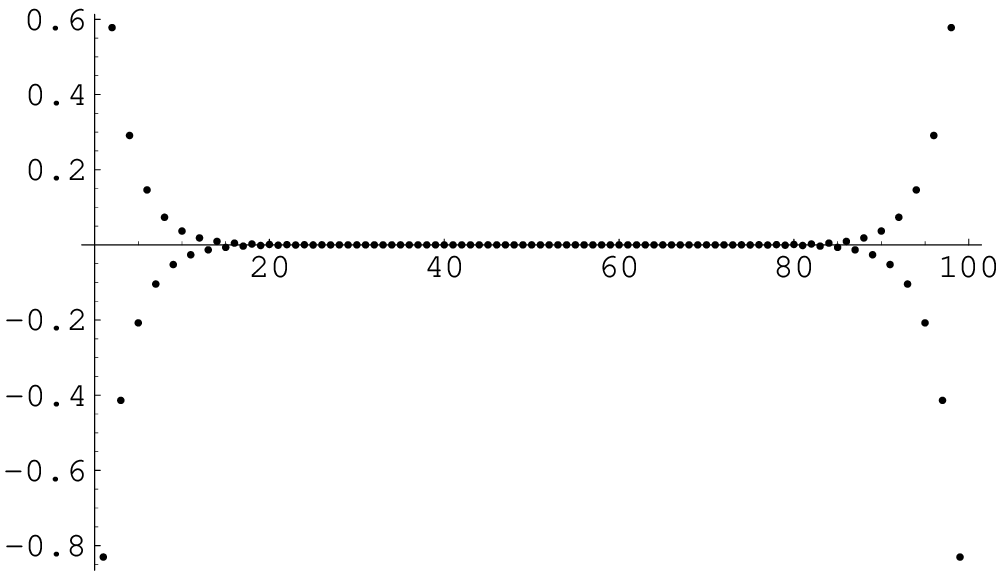, width=7cm}
\hspace{1cm}
\epsfig{file=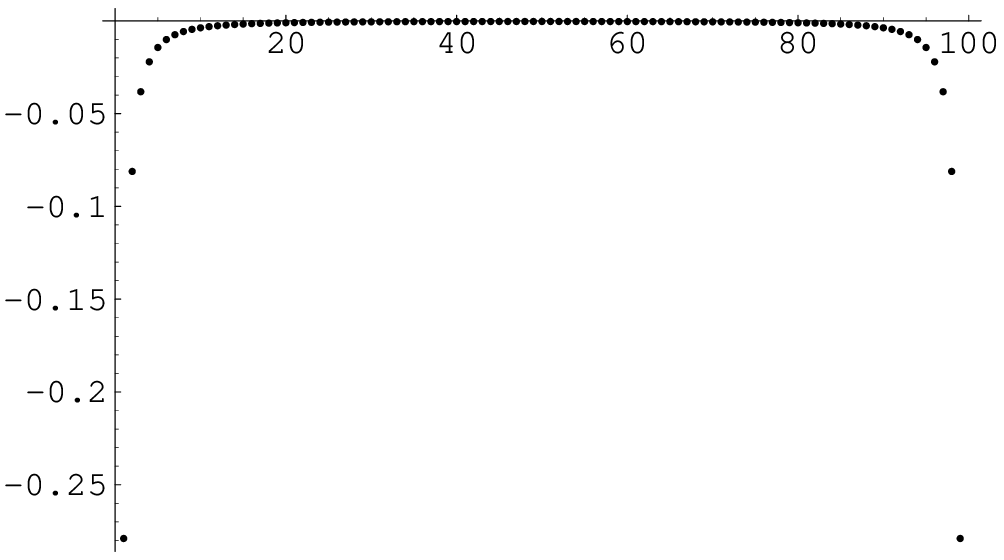, width=7cm}
\end{center}
\vspace*{-1.5cm}
\caption{
The minimum energy non-integer configurations $\{n_r^C\}$ with $n=1$
in the case $N=100$. The value of $b$ is $b=0.1$ in the left figure and
$b=10$ in the right one.
}
\label{fig:n_r^C}
\end{figure}
Figure \ref{fig:n_r^C} shows the configurations $\{n_r^C\}$ in the
cases of $b=0.1$ (left figure) and $b=10$ (right figure) for $N=100$.
As seen from the figure, $\{n_r^C\}$ is localized
around $r=0 \pmod{N}$ to screen the charge $n=1$.\footnote{
As seen from figure \ref{fig:n_r^C}, the charges $n_r^C$ near $r=0$
all have opposite sign to $n_0$ for larger values of $b$, while
$n_r^C$ have alternating signs for smaller $b$.
}

Our second comment is on the term
$\exp\bigl(-\bigl[2\bigl(\wh{Q}^{-1}\bigr)_{00}\bigr]^{-1}
\left(n^2/R^2\right)\bigr)$ in \eqref{tn0Poisson}.
The exponent is equal to the value of the Hamiltonian $H$ for the
configuration $\{n \cdot n_r^C\}$, namely, the (non-integer) configuration
minimizing $H$ for a given $n$:
\begin{equation}
H(n\cdot n_r^C\,;n)=\frac{n^2}{2\bigl(\wh{Q}^{-1}\bigr)_{00}}\,.
\end{equation}
As was analyzed in \cite{RSZ2},
$1/\bigl(\wh{Q}^{-1}\bigr)_{00}$ is finite in the limit
$N\to\infty$.\footnote{
$1/\bigl(\wh{Q}^{-1}\bigr)_{00}$ is related to $S'_{00}$ in
\cite{RSZ2} by
$1/\bigl(\wh{Q}^{-1}\bigr)_{00}
=b\left(1/2 +S'_{00}/(1-S'_{00})\right)
$.
}
\begin{figure}[htbp]
\begin{center}
\leavevmode
\put(205,40){{\large $\bm{\ln b}$}}
\put(55,175){{$\bm{\bigl[2\bigl(\wh{Q}^{-1}\bigr)_{00}\bigr]^{-1}}$}}
\epsfig{file=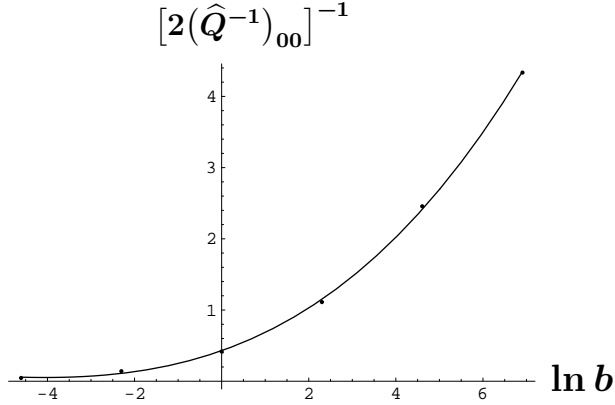, width=7cm}
\end{center}
\vspace{-1.5cm}
\caption{
$\lim_{N\to\infty}\bigl[2\bigl(\wh{Q}^{-1}\bigr)_{00}\bigr]^{-1}$
as a function $\ln b$.
The dots represent
$\lim_{N\to\infty}\bigl[2\bigl(\wh{Q}^{-1}\bigr)_{00}\bigr]^{-1}$
at $b=1/100$, $1/10$, $1$, $10$, $100$, $1000$ obtained by evaluating
its values for $N=50$, $100$, $200$, $300$, $400$, $500$, $600$
and then extrapolating them to $N=\infty$ by using the fitting
function of the form $\sum_{k=0}^3 c_k/N^k$.
The curve interpolating these six points is
$0.429214 + 0.216204\times\ln b + 0.0379729\times(\ln b)^2
+ 0.00186225\times(\ln b)^3$.
}
\label{fig:1/invwhQ00}
\end{figure}
Figure \ref{fig:1/invwhQ00} shows
$\lim_{N\to\infty}
\bigl[2\bigl(\wh{Q}^{-1}\bigr)_{00}\bigr]^{-1}$
as a function of $\ln b$.
It is a monotonically increasing function of $b$ and, as we shall see
in section 4, approaches zero as $b\to 0$.

Now let us consider $t_{n}$ for $R^2\gg 1$.
Since $\hhQ$ is a positive definite matrix, the configuration $m_r=0$
for all $r=1,2,\cdots,N-2$ dominates the $m_r$-summation in
\eqref{tn0Poisson} for $R^2\gg 1$.
Namely, we have
\begin{equation}
t_{n}\simeq
R^{N-2}\Bigl(\det\hhQ\Bigr)^{-1/2}
\exp\!\left(-\frac{1}{
2\bigl(\wh{Q}^{-1}\bigr)_{00}}\frac{n^2}{R^2}\right)\,,
\quad (R^2\gg 1)\,.
\label{tn0R^2gg1}
\end{equation}
This expression can also be obtained by simply replacing the
$n_r$-summations in \eqref{t_n0} with integrations over continuous
variables $p_r=n_r/R$. Since the coefficient of $n^2$ in the exponent
is non-vanishing for $b>0$, this $t_{n}$ cannot lead to a desirable
rolling profile.

\begin{figure}[htbp]
\begin{center}
\leavevmode
\put(205,43){{\large $\bm{n}$}}
\put(-10,167){$\bm{\DltH}$}
\epsfig{file=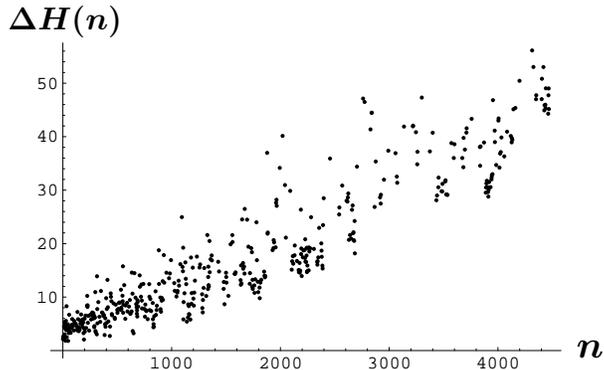, width=7cm}
\end{center}
\vspace{-1.5cm}
\caption{
$\DltH$ v.s.\ $n$ for $N=2048$ and $b=0.1$. For obtaining $\DltH$, we
approximate the integer-valued charge $n_r^{\rm I}(n)$ by the integer
nearest to $n\cdot n_r^{\rm C}$. However, this $\{n_r^{\rm I}(n)\}$
does not necessarily satisfy the constraint
$n+\sum_{r=1}^{N-1}n_r^{\rm I}(n)=0$.
In the figure, only the points which satisfy the constraint are
plotted. Distribution of the points is insensitive to the value of
$N$ if it is large enough.
}
\label{fig_deltaH}
\end{figure}

On the other hand, $t_{n}$ \eqref{t_n0} for $R^2\ll 1$ can be
approximated by
\begin{equation}
t_{n}\simeq \exp\!\left(
-\frac{1}{R^2}H\!\left(n_r^I(n);n\right)\right)\,,
\quad (R^2\ll 1)\,,
\label{tn0R^2ll1}
\end{equation}
where $\{n_r^I(n)\}$ is the {\em integer} configuration which minimizes
the Hamiltonian for a given $n$.
This configuration $\{n_r^I(n)\}$ is in general different from but
is close to $\{n\cdot n_r^C\}$, the configuration minimizing $H(n_r;n)$
without the integer restriction. Therefore, rewriting
\eqref{tn0R^2ll1} as
\begin{equation}
t_{n}\simeq \exp\!\left(
-\frac{1}{
2\bigl(\wh{Q}^{-1}\bigr)_{00}}\frac{n^2}{R^2}\right)
\times
\exp\!\left(-\frac{1}{R^2}\,\DltH\right)\,,
\label{tn0R^2ll1_2}
\end{equation}
with $\DltH$ defined by
\be
\DltH=H\!\left(n_r^I(n);n\right)-H\!\left(n\cdot n_r^C;n\right)\,,
\ee
the second factor of \eqref{tn0R^2ll1_2} is expected to have a milder
$n$-dependence than the first one.
Figure \ref{fig_deltaH} shows $\DltH$ for $N=2048$ and $b=0.1$.
We see that $\DltH$ is in fact roughly proportional to $n$.
Therefore, we cannot obtain a rolling profile in the case $R^2\ll 1$
either.

Summarizing this subsection, for both $R^2\gg 1$ and $R^2\ll 1$,
the coefficient $t_n$ of the component field $t(x)$ has the leading
$n$-dependence of the form
\be
t_{n}\sim \exp\left(-\frac{1}{2(\wh{Q}^{-1})_{00}}\,\frac{n^2}{R^2}\,
\right)\,.
\label{tnallR^2}
\ee
Then, $t(x)$ itself shows the behavior
\begin{align}
t(x)&\sim
\exp\left(\frac{1}{2}(\wh{Q}^{-1})_{00}(x)^2 \right)
\times ({\mbox{oscillating part}})
\,,
\label{tx0}
\end{align}
and cannot approach the tachyon vacuum as $x\to\infty$. This is the
case even if we adopt the alternating sign solution $(-1)^n t_n$.

\subsection{Numerical analysis using Monte Carlo simulation}

For studying $t_{n}$ and $t(x)$ for intermediate values of $R^2$, we
shall carry out Monte Carlo simulation of the Coulomb system with
partition function \eqref{tn}, Hamiltonian $H$ and temperature $R^2$.
We have adopted the Metropolis algorithm.
Since the total charge must be kept zero, a new configuration is
generated from the old one $\{n_r\}$ by randomly choosing two points
$r$ and $s$ on the circle and making the change
$(n_r,n_s)\to(n_r+1,n_s-1)$. This new configuration is
accepted/rejected according to the standard Metropolis algorithm.

\subsubsection{Time derivatives of $\bm{\ln t(x)}$}
\label{MonteCarlot(x)}

First we investigate the time derivatives of the logarithm of
$t(x)$. Defining $T(x)$ by
\be
t(x)=e^{T(x)}\, ,
\ee
we have
\begin{align}
\frac{d T(x)}{d x}&=\frac{1}{R}\,\VEV{n}_{x}\, ,
\label{vel}
\\[2mm]
\frac{d^2 T(x)}{d {x}^2}&=\frac{1}{R^2}
\left(\VEV{n^2}_x -\VEV{n}_{x}^2\right) \, ,
\label{accel}
\end{align}
where the average $\VEV{\cO}_x$ for a given $x$ is defined by
\be
\VEV{\cO}_x=\frac{1}{t(x)}\sum_{
\begin{smallmatrix}
n,n_1,\cdots,n_{N-1}=-\infty\\
\left(n+\sum_{r=1}^{N-1}n_r=0\right)
\end{smallmatrix}
}^\infty \!\!\cO\,e^{-H(n_r;n)/R^2 +nx/R} \,.
\ee

\begin{figure}[htbp]
\vspace{-0.2cm}
\begin{center}
\leavevmode
\put(205,44){{\large$\bm{x}$}}
\put(-5,169){$\ds\bm{dT(x)/dx}$}
\epsfig{file=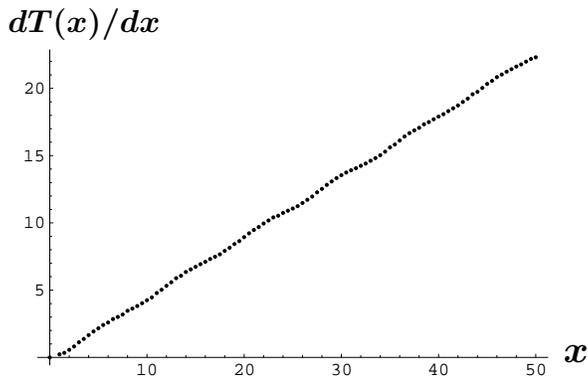, width=7.0cm}
\vspace{-1.3cm}
\caption{The numerical results of $dT(x)/dx$
at $x=0.5,1.0,1.5,\cdots,50.0$. Here we have taken $b=10$, $R^2=1$ and
$N=256$. These points are well fitted by a linear function
$0.4530 x -0.1713$, and the slope $0.4530$ is close to
$(\hat{Q}^{-1})_{00}=0.4497$ for the present $b$ and $N$.
The small oscillatory behavior can be better observed in
$d^2T(x)/dx^2$ shown in figure \ref{fig_n02}.
}
\label{fig_n0}
\end{center}
\end{figure}
\begin{table}[htbp]
\begin{center}
\begin{tabular}{|c|c|c|c|c|c|c|c||c|}
\hline
$R^2$ &  0.3&0.5&1.0 &1.5&2.0&3.0&5.0 & $(\wh{Q}^{-1})_{00}$\\ \hline
slope & $0.4557$ & $0.4540$ & $0.4530$ & $0.4502$ & $0.4498$ &
 $0.4498$ & $0.4497$ & $0.4497$ \\\hline
\end{tabular}
\end{center}
\caption{The slope of the linear function of $x$ obtained by fitting
the Monte Carlo results of $\VEV{n}_{x}/R$ with $b=10$ and $N=256$.
They are almost independent of $R^2$ and close to
$(\wh{Q}^{-1})_{00}$.
}
\label{coefficient}
\end{table}
The numerical results of the ``velocity'' \eqref{vel} versus $x$
for $b=10$, $R^2=1$ and $N=256$ are shown in figure \ref{fig_n0}.
We find that $dT(x)/dx$ is almost linear in $x$.
The slope of the fitted linear function for the various $R^2$ and
$b=10$ and $N=256$ as well as the value of $(\wh{Q}^{-1})_{00}$ for
the present $b$ and $N$ are shown in table \ref{coefficient}.
These results show that the behavior of \eqref{tx0}
obtained in the regions $R^2\ll 1$ and $R^2\gg 1$ is valid also in
the intermediate region of $R^2$.

\begin{figure}[htbp]
\vspace{-0.5cm}
\begin{center}
\leavevmode
\put(247,72){{\large$\bm{x}$}}
\put(240,132){$\bm{(\wh{Q}^{-1})_{00}}$}
\put(0,199){$\ds\bm{d^2 T(x)/dx^2}$}
\epsfig{file=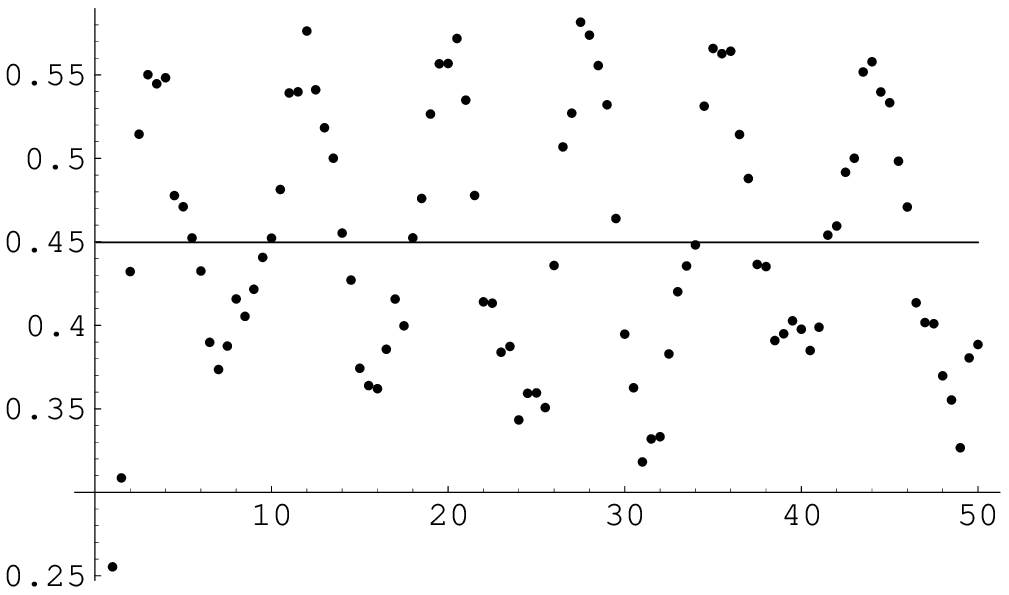, width=8.5cm}
\vspace{-1.5cm}
\caption{$d^2 T(x)/dx^2$ at $x=0.5,1.0,1.5,\cdots,50.0$. Here we have
taken $b=10$, $R^2=1$ and $N=256$. The horizontal line shows the value
of $(\wh{Q}^{-1})_{00}=0.4497$ for the present $b$ and $N$.
}
\label{fig_n02}
\end{center}
\end{figure}
The Monte Carlo results of the ``acceleration'' \eqref{accel}
at various $x$ are shown in figure \ref{fig_n02} in the case of
$b=10$, $R^2=1$ and $N=256$.
The acceleration oscillates with period roughly equal to $8$.
The center of the oscillation is around $(\wh{Q}^{-1})_{00}=0.4497$,
which is consistent with \eqref{tx0}.\footnote{
In this case, the parameter $a$ of \eqref{tn0behavior} is
$a=\bigl[2(\wh{Q}^{-1})_{00}R^2\bigr]^{-1}=1.112$.
If the subleading part $\wt{t}_n$ of \eqref{tn0behavior} is
independent of $n$, the oscillation period of $t(x)$ should be
$2a=2.224$. The fact that the period of figure \ref{fig_n02} is nearly
equal $8$, which is four times the naive period, suggests that
$\wt{t}_{4k+1}$, $\wt{t}_{4k+2}$ and $\wt{t}_{4k+3}$ are negligibly
small compared with $\wt{t}_{4k}$.
}
We have studied the acceleration for other values of $R^2$ and
found that it oscillates around $(\wh{Q}^{-1})_{00}$ for any $R^2$.

\subsubsection{Numerical analysis of $\bm{t_{n}}$}
\label{Numanalysistn}

The $n$-dependence of $t_n$ itself can be directly measured using
Monte Carlo simulation as follows \cite{Fukaya}.
Here, we use instead of $R^2$ the inverse temperature $\beta=1/R^2$,
and make explicit the $\beta$-dependence of $t_n$ to write it as
$t_{n,\beta}$. Let us define the average $\VEV{\cO}_{n,\beta}$ with
subscript $n$ and $\beta$ by
\be
\VEV{\cO}_{n,\beta}=\frac{1}{t_{n,\beta}}\sum_{
\begin{smallmatrix}
n_1,\cdots,n_{N-1}=-\infty\\
\left(n+\sum_{r=1}^{N-1}n_r=0\right)
\end{smallmatrix}
}^\infty \cO\,e^{-\beta H(n_r;n)}\, .
\ee
Integrating the relation
\be
\Drv{}{\beta}\ln t_{n,\beta}=\VEV{-H\,}_{n,\beta}\, ,
\ee
with respect to $\beta$, we obtain
\be
\frac{t_{n,\beta}}{t_{n,\beta=0}}=\exp\left(
\int_0^{\beta}d\beta' \VEV{-H\,}_{n,\beta'}\right)\,,
\ee
and hence
\be
\frac{t_{n,\beta}}{t_{n=0,\beta}}=
\frac{t_{n,\beta=0}}{t_{n=0,\beta=0}}\exp\left(
\int_0^{\beta}d\beta'\left(\VEV{-H\,}_{n,\beta'}
-\VEV{-H\,}_{n=0,\beta'}\right)\right)\,.
\ee
Eq.\ \eqref{tn0R^2gg1} implies that $t_{n,\beta=0}$ is independent of
$n$ and therefore $t_{n,\beta=0}/t_{n=0,\beta=0}=1$.
Thus, we obtain the formula
\be
\frac{t_{n,\beta}}{t_{n=0,\beta}}=
\exp\left(-
\int_0^{\beta}d\beta' \bigr(\VEV{H}_{n,\beta'}
-\VEV{H}_{n=0,\beta'}\bigr)\right)\,.
\label{fukayaformula}
\ee
This allows us, in principle, to directly evaluate the $n$-dependence
of $t_n$ using the expectation values of $H$ obtained by Monte Carlo
simulation.
Note that $\VEV{H}_{n,\beta}$ in high temperature region $\beta\ll 1$
is given using \eqref{tn0R^2gg1} by
\be
\VEV{H}_{n,\beta}\simeq
\frac{N-2}{2\beta}+\frac{1}{2(\wh{Q}^{-1})_{00}}\cdot n^2\,,
\quad (\beta\ll 1)\, .
\label{VEVH_high}
\ee

\begin{figure}[htbp]
\begin{center}
\leavevmode
\put(20,200){$\bm{\VEV{H}_{n,\beta}
-\VEV{H}_{n=0,\beta}-n^2/(2(\wh{Q}^{-1})_{00})}$}
\put(232,47){{\large$\bm{\beta}$}}
\put(230,170){$\bm{n\!=\!100}$}
\put(230,125){$\bm{n\!=\!40}$}
\put(230,105){$\bm{n\!=\!20}$}
\put(230,78){$\bm{n\!=\!10}$}
\epsfig{file=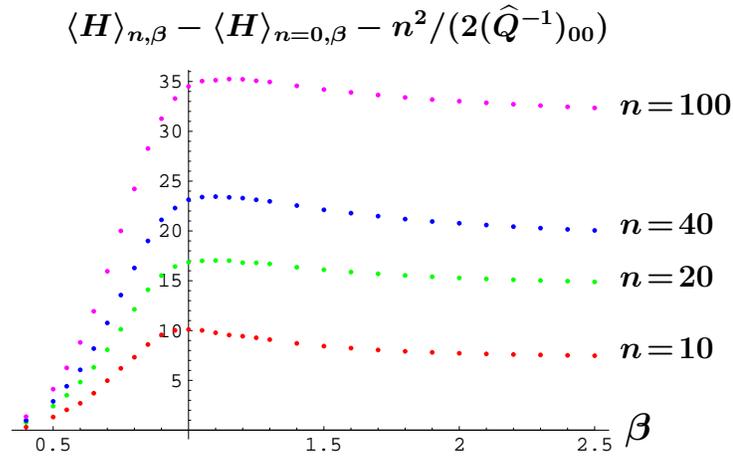, width=8.0cm}
\vspace{-1.3cm}
\caption{
$\VEV{H}_{n,\beta}-\VEV{H}_{n=0,\beta}-n^2/(2(\wh{Q}^{-1})_{00})$
for $n=10$ (red points), $n=20$ (green), $n=40$ (blue), and $n=100$
(purple). Here we have taken $N=256$ and $b=10$.
}
\label{fig:Hn-Hn0_b10}
\end{center}
\end{figure}
Figure \ref{fig:Hn-Hn0_b10} shows the Monte Carlo result of
$\VEV{H}_{n,\beta}-\VEV{H}_{n=0,\beta}-n^2/(2(\wh{Q}^{-1})_{00})$,
namely, the deviation of $\VEV{H}_{n,\beta}-\VEV{H}_{n=0,\beta}$ from
the high temperature value, for the various values of $n$
($b=10$ and $N=256$).
As $\beta$ is decreased, the data for each $n$ approach zero.
On the other hand, they seem to approach a positive constant as
$\beta$ is increased.
Since the asymptotic value at large $\beta$ grows no faster than
linearly in $n$, the results of figure \ref{fig:Hn-Hn0_b10} together
with the formula \eqref{fukayaformula} seems consistent with our low
temperature analysis using \eqref{tn0R^2ll1_2} and figure
\ref{fig_deltaH}.

{}From our analyses in this subsection we have found that the
behaviors \eqref{tnallR^2} of $t_n$ and \eqref{tx0} of $t(x)$ in the
$R^2\gg 1$ and $R^2\ll 1$ regions are valid for any $R^2$.
Concerning these behaviors, $R=1$ does not seem to be a special
radius.

\section{Possible rolling solution with $\bm{b=0}$}
\label{solution_b0}

Our analysis in the previous section implies that rolling solutions
with desirable profile can never be obtained unless the coefficient of
$n^2$ in the exponent of \eqref{tnallR^2} vanishes. Namely, we must
have
\be
\frac{1}{(\wh{Q}^{-1})_{00}}=0\,.
\label{Qcond}
\ee
This condition can in fact be realized by putting $b=0$ as can be
inferred from figure \ref{fig:1/invwhQ00}, though
the precise way how this condition is satisfied differs between the
$N=\mbox{even}$ and the odd cases.
For an even (and finite) $N$, the condition \eqref{Qcond} is satisfied
by simply putting $b=0$.
This is because $\wh{Q}$ with $N=\mbox{even}$ and $b=0$ has a
zero-mode $\{(-1)^r\}$:
\be
\sum_{s=0}^{N-2}\wh{Q}_{rs}(-1)^s=0\,,
\quad(r=0,1,\cdots,N-2;\, N=\mbox{even}, b=0)\, .
\label{hatQ(-1)^s=0}
\ee
This zero-mode is at the same time the minimum energy configuration
$n_r^C$ \eqref{nrC}.\footnote{
Note that the condition for the minimum energy configuration is
$\sum_{s=0}^{N-2}\wh{Q}_{rs}n_r^C=0$ for $r=1,2,\cdots,N-2$, while the
zero-mode of $\wh{Q}$ should satisfy this equation for
$r=0,1,\cdots,N-2$ including $r=0$.
}

On the other hand, in the $N=\mbox{odd}$ case, the condition
\eqref{Qcond} is realized by putting $b=0$ and in addition taking the
limit $N\to\infty$. In fact, numerical analysis shows that
\be
\frac{1}{(\wh{Q}^{-1})_{00}\bigr|_{b=0}}
=\frac{1.705}{N}+O\!\left(\frac{1}{N^2}\right)\, ,
\quad (N\gg 1, N=\mbox{odd})\, .
\ee
This property is related to the fact that the matrix $\wh{Q}$ with $b=0$
has an approximate zero-mode for large and odd $N$.
This approximate zero-mode, which is at the same time the
minimum energy configuration $n_r^C$ \eqref{nrC}, is given by
\be
n_r^C\simeq (-1)^r\times \left(1-\frac{2\abs{r}}{N+1}\right)\, ,
\quad
\left(r=0,\pm 1,\pm 2,\cdots,\pm(N-1)/2\right)\, ,
\label{zero_odd}
\ee
where the index $r$ should be understood to be defined
$\mbox{mod}\,N$; $n_r=n_{r+N}$.
This configuration satisfies
$\sum_{r=0}^{N-1}n_r^C=0$ and
\be
\sum_{s=0}^{N-2}\wh{Q}_{rs}n_s^C
=O\!\left(\frac{1}{N}\right)\,,\quad
(r=0,1,\cdots,N-2;\,N\gg 1, b=0)\, .
\label{whQnC=1/N}
\ee
A proof of \eqref{whQnC=1/N} is given in the appendix.

In the rest of this section we shall consider only the case of
$N=\mbox{odd}$ since \eqref{prop_wedge} is closed among the
$\ket{N=\mbox{odd}}$ states ($\ket{N}$ is the state given by
\eqref{compsliver} in the present case).
Here we shall just make a comment on the case of $N=\mbox{even}$.
When $N=\mbox{even}$ and $b=0$, the coefficient
$t_n$ is {\em independent} of $n$ as can be seen from the expression
\eqref{tn0Poisson} with $n_r^C=(-1)^r$. Therefore, we have
$t(x)=t_0 \sum_{n=-\infty}^\infty (\pm 1)^n e^{nx/R}$.
A naive summation of this series gives $t(x)=0$, or
the Poisson's resummation formula gives
$t(x)=2\pi t_0\sum_{m={\rm even}/{\rm odd}}
\delta\left(ix/R-\pi m\right)$
\cite{imaginary}.

We would like to investigate whether the solution with $b=0$ and
$N(={\rm odd})\to\infty$ can be regarded as a rolling solution.
However, it is not an easy matter to repeat the Monte Carlo analysis
of Sections \ref{MonteCarlot(x)} and \ref{Numanalysistn} in the
present case. First, since the coefficient $(\wh{Q}^{-1})_{00}$ of the
leading $x$-dependent term of \eqref{tx0} blows up as
$N\to\infty$,\footnote{
This divergence is only an apparent one coming from applying the
evaluation of $t(x)$ given in \eqref{evalt(x)} to the case with an
infinitesimally small $a$.
If the leading term \eqref{tnallR^2} of $t_n$ is missing
from the start, we have to adopt a different way of estimating the
summation \eqref{tx}.
}
so do the slope of the velocity curve of figure \ref{fig_n0} and the
central value of the acceleration curve of figure \ref{fig_n02}.
Therefore, it is hard to read off the subleading $x$-dependence which
should become the leading one in the limit $N\to\infty$.
Second, the direct evaluation of the coefficient $t_n$ using the
formula \eqref{fukayaformula} is not an easy task because of bad
statistics problem. Namely, the difference
$\VEV{H}_{n,\beta}-\VEV{H}_{n=0,\beta}$, which is of order
$n^2/(2(\wh{Q}^{-1})_{00})$ (see \eqref{VEVH_high}), is very small
compared with the leading bulk term $(N-2)/(2\beta)$ of $\VEV{H}$ when
$b=0$.

\begin{figure}[htbp]
\begin{center}
\leavevmode
\put(248,57){{\large$\bm{n}$}}
\put(0,207){$\bm{\DltH}$}
\epsfig{file=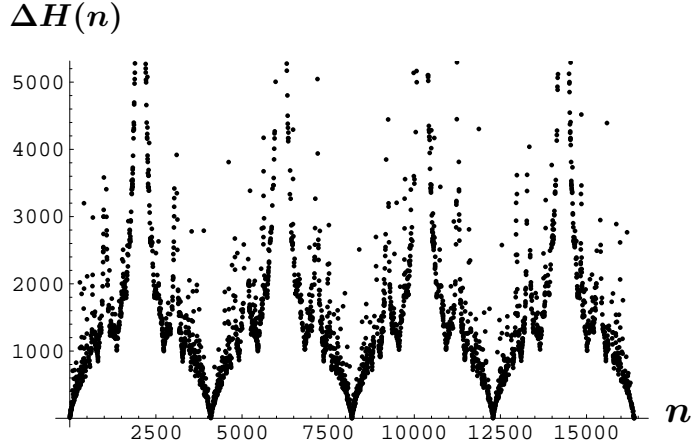, width=8.5cm}
\vspace{-1.5cm}
\caption{$\DltH$ v.s. $n$
for $N=8191$ and $b=0$.
}
\label{DHN8191}
\end{center}
\end{figure}

Here we shall content ourselves with the analysis of $t_n$ in the low
temperature region $R^2\ll 1$ using the expression
\eqref{tn0R^2ll1_2}.
Since the leading term
$\exp\bigl(-n^2/(2(\wh{Q}^{-1})_{00}R^2)\bigr)$
of \eqref{tn0R^2ll1_2} disappears in the limit $N\to\infty$,
we study the $n$-dependence of $\DltH$ in the same way as we did
in figure \ref{fig_deltaH}.
The result for $N=8191$ is plotted in figure
\ref{DHN8191}.

Note that the data of figure \ref{DHN8191} has an approximate periodic
structure with respect to $n$ with period of about $4100$.
This periodicity can be understood from \eqref{tn0Poisson} for $t_n$
and the expression \eqref{zero_odd} for $n_r^C$ independently of the
low temperature approximation.
In fact, \eqref{tn0Poisson} without the leading term
$\exp\bigl(-n^2/(2(\wh{Q}^{-1})_{00}R^2)\bigr)$
is invariant under the shift $n\to
n+(N+1)/2$ since the change of the exponent in the $m_r$-summations
under this shift is an integer multiple of $2\pi i$ for $n_r^C$ of
\eqref{zero_odd}.
Although figure \ref{DHN8191} shows data only for positive values of
$n$, recall that $t_n$ is even under $n\to -n$; \eqref{tniseven}.
This parity symmetry and the periodicity lead to the structure shown
in figure \ref{DHN8191}.

The periodicity stated above,
\be
t_{n+(N+1)/2}=t_n\, ,
\label{periodicity}
\ee
is not an exact one for a finite $N$ since both the condition
\eqref{Qcond} and eq.\ \eqref{zero_odd} for $n_r^C$ are only
approximately satisfied.
If the periodicity \eqref{periodicity} were exact, $t(x)$ given by
\eqref{tx} could be rewritten as
\be
t(x)= \tpart(x)\sum_{k=-\infty}^{\infty}
\exp\!\left(\frac{N+1}{2R}\,kx\right)\, ,
\label{t=tpartsum}
\ee
where $\tpart(x)$ is defined by
\be
\tpart(x)=\sum_{n=-[N/4]}^{[(N+1)/4]}t_n\,e^{n x/R}\, ,
\label{tpart}
\ee
with $[c]$ being the largest integer not exceeding $c$.
Namely, $\tpart(x)$ is the one period part around $n=0$ in the
summation of $t(x)$.
The geometric series multiplying \eqref{t=tpartsum} is formally summed
up to give an unwelcome result; it is equal to zero or the summation
of delta functions for pure imaginary values of $x$.
However, since the periodicity \eqref{periodicity} is not exact for a
finite $N$, we have to carry out more precise analysis taking into
account the violation of the periodicity to obtain the profile $t(x)$
in the limit $N\to\infty$.
Here we would like to propose another way of defining $t(x)$ which
could lead to a desirable rolling profile.
It is the $N\to\infty$ limit of the one period summation
\eqref{tpart}:
\be
t(x)=\lim_{N\to\infty}\tpart(x)\, .
\label{t(x)_2}
\ee
This is also formally equal to the original summation \eqref{tx}
with $N=\infty$.

\begin{figure}[htbp]
\begin{center}
\leavevmode
\put(233,53){{\large$\bm{n}$}}
\put(0,188){$\bm{\DltH}$}
\epsfig{file=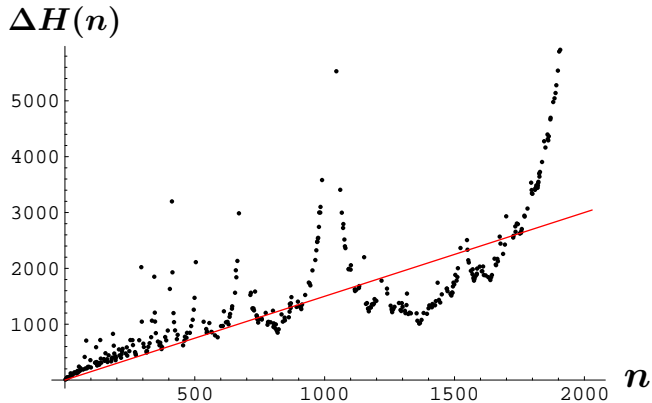, width=8cm}
\vspace{-1.5cm}
\caption{
$\DltH$ for $n\in[0,2048]$ in the case $N=8191$.
The red line is an auxiliary one with slope 3/2.
}
\label{DHN8191N/4}
\end{center}
\end{figure}

Let us return to the low temperature analysis with $R^2\ll 1$.
For $t(x)$ given by \eqref{t(x)_2}, it is sufficient to study the
$n$-dependence of $\DltH$ only in the half period region
$n\in\bigl[0,[(N+1)/4]\bigr]$, which is shown in figure
\ref{DHN8191N/4} for $N=8191$.
As shown in figure \ref{DHN8191N/4}, $\DltH$ has a complicated
structure with peaks and valleys. However, if we neglect such local
structures and see figure \ref{DHN8191N/4} globally, we find that
$\DltH$ grows almost linearly in $n$; $\DltH\propto n$
(the red line in figure \ref{DHN8191N/4} is the line with
slope $3/2$).
This could give a desirable $t_n$ with the behavior \eqref{tn=beta^n}
up to complicated local structures.
However, it is a nontrivial problem whether the $N\to\infty$ limit of
\eqref{t(x)_2} really exists even when we take into account the local
structures.
Here we shall point out a kind of self-similarity of $\DltH$ and hence
of $t_n$; $2\,\DltH|_{N}\simeq \Delta H(2n)|_{2N}$.
\begin{figure}[htbp]
\vspace{-1.0cm}
\begin{center}
\leavevmode
\put(233,53){{\large$\bm{n}$}}
\put(0,188){$\bm{\DltH}$}
\epsfig{file=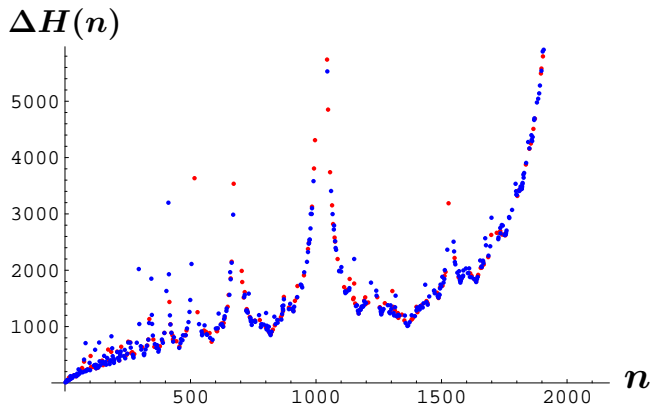, width=8cm}
\vspace{-1.5cm}
\caption{
$\DltH$ for $N=8191$ (blue points) and that for $N=4095$ (red
points). Both the horizontal and the vertical scales are doubled for
the red points.
}
\label{DH2}
\end{center}
\end{figure}
Figure \ref{DH2} shows
$\DltH$ for $N=8191$ (blue points) and that for $N=4095$ (red
points). Both the horizontal and the vertical scales are doubled for
the $N=4095$ points. For example, the real coordinate of the red point
$(1500,1600)$ in the figure is actually $(750,800)$.
Note that the red points and the blue ones have overlapping local
structures. It is our future subject to study whether $t(x)$ of
\eqref{t(x)_2} can exist for $t_n$ with such self-similarity.

\section{Thermodynamic properties of $\bm{t(x)}$}

In the BCFT analysis, the rolling tachyon solution is obtained by the
inverse Wick rotation of one space direction compactified on a circle
at the self dual radius $R=1$ \cite{marginal,rolling}.
Therefore, also in our VSFT construction of the rolling tachyon
solution, it is natural to expect that the meaningful solution exists
only at $R=1$.
This is also supported by the fact that
the correct tachyon mass squared $-1$ has been successfully reproduced
in the analysis of the fluctuation modes around the D25-brane solution
in VSFT \cite{HataVSFT,RSZnote,Okawa}.
If the tachyon mass squared is equal to $-1$, the natural mode of
the expansion in \eqref{tx} is $e^{n x}$ with $R=1$
\cite{padic,Fujita}; in particular, the $n=\pm 1$ modes $e^{\pm x}$
are the massless modes at the unstable vacuum.

In this section we shall study how this critical radius $R=1$
appears in our construction of time-dependent solution in VSFT,
especially in the component field $t(x)$.
One would naively expect that the $N\to\infty$ limit of our solution
\eqref{Psi} can exist only at $R=1$.
Here, we do not pursue this possibility directly, but instead
address the problem from a statistical mechanics point of view.
Recall that $t(x)$ \eqref{tx} and $t_n$ \eqref{tn} can be interpreted
as the partition functions of a statistical system of charges
located on a unit circle with temperature $R^2$.
One possible mechanism of $R^2=1$ being a special point for this
statistical system is that it undergoes some kind of phase transition
at $R^2=1$.
In this section, we first claim, on the basis of a simple
energy-entropy argument, the presence of a phase transition at
$R^2=1$. Then, we study in more detail the thermodynamic properties of
the system both analytically and numerically.
Our results here suggest but do not definitely confirm the presence
a phase transition at $R^2=1$.

\subsection{Boundstate phase and dissociated state phase}
\label{BSandDS}

Let us consider the statistical system with partition function
$t(x=0)$:\footnote{
Although we consider here $t(0)$ with $x=0$, $t(x\neq 0)$ and $t_n$
have the same bulk thermodynamic properties since the difference is
only the local one at $r=0$.
}
\be
t(0)=\!\!\!\sum_{
\begin{smallmatrix}
n_0,n_1,\cdots,n_{N-1}=-\infty\\
(n_0+n_1+\cdots +n_{N-1}=0)
\end{smallmatrix}}^\infty
\!\!\!\exp\!\left(-\frac{1}{R^2} H(n_r;n_0)\right)\, .
\label{t(0)}
\ee
We would like to argue that this system has a possible phase
transition at temperature $R^2=1$.
In the low temperature region $R^2\ll 1$, configurations with lower
energy contribute more to the partition function.
The lowest energy configuration of the Hamiltonian $H$ \eqref{H} is
of course that with all $n_r=0$.
Due to the self-energy part $2\ln(N/4)$ of $Q_{rr}$ \eqref{Qrs},
the energy of a generic configuration with zero total charge can be
$\ln N$-divergent. Finite energy configurations are those where the
charges are confined in finite size regions and the sum of charges in
each region is equal to zero. Namely, they consist of neutral
boundstates of charges.
The simplest among them is the configuration
of a pair of $+1$ and $-1$ charges with a finite separation.
Let us consider the configuration $\{n_r^{(k,\Delta)}\}$
with $n_k=+1$, $n_{k+\Delta}=-1$ and all other $n_r=0$ for a given
position $k$ and separation $\Delta$. The energy of this configuration
is
\be
H\left(\{n_r^{(k,\Delta)}\}\right)
=\frac{b}{2}+2\ln\!\left(\frac{N}{4}\right)
+2\ln\abs{2\sin\left(\frac{\pi\Delta}{N}\right)}\, .
\label{energy}
\ee
This is approximated in the close case $\Delta=O(1)\pmod N$ and in
the far separated case $\Delta=O(N)\pmod N$ by
\be
H\left(\{n_r^{(k,\Delta)}\}\right)\sim
\begin{cases}
\ds \frac{b}{2}+2\ln\left(\frac{\pi\abs{\Delta}}{2}\right)\,,
& \Delta=O(1) \,,\\
\ds 2\ln N \,,&\Delta=O(N) \,.
\end{cases}
\label{Hpair}
\ee
The energy in the finite separation case $\Delta=O(1)$ is indeed free
from the $\ln N$ divergence.
Other neutral boundstates with finite energy are, for example,
a pair of charges $+q$ and $-q$ with $q>1$, and a chain of alternating
charges
\be
(0,0\cdots,0,q,-q,q,-q,\cdots,q,-q,0,\cdots,0)\, .
\label{chain}
\ee

Configurations with isolated charges have $\ln N$-divergent energy as
seen from the $\Delta=O(N)$ case of \eqref{Hpair}.
However, this does not imply that such configurations do not
contribute at all to the partition function: we have to take into
account their entropy. Let us consider a naive free energy argument of
an isolated charge.
As seen from $Q_{rs}$ \eqref{Qrs} or \eqref{Hpair} for $\Delta=O(N)$,
the energy of an isolated charge is ${\cal E}=\ln N$, while the
entropy of this charge is ${\cal S}=\ln N$ since there are $N$ points
where it can sit on.
Therefore the free energy of this isolated charge is given by
\be
{\cal F}_{\mbox{\scriptsize isolated charge}}={\cal E}-R^2{\cal S}
=(1-R^2)\ln N \, .
\label{Fisolatedcharge}
\ee
This means that, for $R^2>1$, the free energy becomes lower as more
isolated charges are excited. Namely, $R^2=1$ could be a phase
transition point separating the boundstate phase in $R^2<1$ and
the dissociated state phase in $R^2>1$.
To confirm the existence of this phase transition, more precise
analysis is of course necessary.

\subsection{Dilute pair approximation}
\label{secDPA}

Before carrying out numerical studies of the system \eqref{t(0)} for
the possible phase transition at $R^2=1$, we shall in this subsection
present some analytic results valid in low temperature.
In the low temperature region $R^2\ll 1$, it should be a good
approximation to take into account only the pairs of charges as
configurations contributing to the partition function \eqref{t(0)}.
This approximation is better for larger $b$ since more complicated
boundstates such as \eqref{chain} have larger energy coming from the
$b/2$ term of $Q_{rr}$ \eqref{Qrs}.
The partition function of a pair of charges is
given by summing over the position $k$ and the separation $\Delta$
of the configuration $\{n_r^{(k,\Delta)}\}$:
\be
\Zop=\sum_{k=0}^{N-1}\sum_{\Delta=1}^{N-1}
e^{-H(\{n_r^{(k,\Delta)}\})/R^2}\, .
\label{Zop}
\ee
In the low temperature region where the number of pair excitations is
small and the pairs are far separated from each other, we can
exponentiate $\Zop$ to obtain
\be
t(0)=\exp\left(\Zop\right)\,.
\label{Z_pairs}
\ee
We call this ``dilute pair approximation''.
Using \eqref{energy}, $\Zop$ is calculated as follows:
\begin{align}
\Zop&=\sum_{k=0}^{N-1}\sum_{\Delta=1}^{N-1} e^{-b/(2R^2)}
\left(\frac{N}{2}\right)^{-2/R^2}
\abs{\,\sin\frac{\pi\Delta}{N}}^{-2/R^2}\nn\\
&=e^{-b/(2R^2)}\left(\frac{N}{2}\right)^{-2/R^2}
\frac{N^2}{\pi}\int_{\pi/N}^{\pi(1-1/N)}dy
\left( \sin y\right)^{-2/R^2}
\nn\\
&=
\begin{cases}
\ds \frac{2N}{2/R^2-1}
\left(\frac{\pi e^{b/4}}{2}\right)^{-2/R^2}\,,
& \left(R^2<2\right) \,,
\\[12pt]
\ds N^{2-2/R^2} \left(\frac{e^{b/4}}{2}\right)^{-2/R^2}
\frac{\Gamma \bigr(\frac12-\frac{1}{R^2}\bigr)}{\sqrt{\pi}\,
\Gamma\bigr(1-\frac{1}{R^2}\bigr)}\,, &  \left(R^2>2\right) \,.
\end{cases}
\label{molecule}
\end{align}
In the second line of \eqref{molecule} we changed the
$\Delta$-summation to the integral with respect to $y=\pi\Delta/N$ for
$N\gg 1$.
This integral is divergent (convergent) at the edges as $N\to\infty$
in the region $R^2<2$ ($R^2>2$).
The final result of \eqref{molecule} in $R^2<2$ has been obtained by
taking the contribution near the edges $y\sim\pi/N$ and $\pi(1-1/N)$,
while that in $R^2>2$ by extending the $y$-integration region to
$[0,\pi]$.

As we raise $R^2$, the more pairs are excited and the separation
of the two charges of a pair becomes larger, leading to the breakdown
of the dilute pair approximation.
{}From \eqref{molecule}, one might think that this breakdown occurs at
$R^2=2$.
However, let us estimate more precisely the limiting temperature above
which the dilute pair approximation is no longer valid.
The criterion for the validity of the dilute pair approximation is
given by
\be
\VEV{p}\times\overline{\Delta}
\lesssim N \, ,
\label{criterion}
\ee
where $\VEV{p}$ and $\overline{\Delta}$ are the average number of
pairs and the average separation between $+1$ and the $-1$ charges of a
pair, respectively.  First, $\VEV{p}$ is given by
\be
\VEV{p}=\frac{1}{t(0)}\sum_{p=0}^\infty\frac{p}{p!}(\Zop)^p
=\Zop\, .
\label{VEVp}
\ee
The average separation $\overline{\Delta}$ is calculated as follows:
\begin{align}
\overline{\Delta}&=
\frac{N}{\pi}\times
\frac{\ds \int_{\pi/N}^{\pi(1-1/N)}\!\!\!\!dy\,
\min(y,\pi-y)\,(\sin y)^{-2/R^2}
}{\ds \int_{\pi/N}^{\pi(1-1/N)}\!\!\!\!dy\,(\sin y)^{-2/R^2}}
\nn\\
&=
\begin{cases}
\ds \frac{1-R^2/2}{1-R^2}\, , & (R^2 <1)\, ,\\[10pt]
\ds \left(\frac{N}{\pi}\right)^{2-2/R^2}
\left(\frac{1}{R^2}-\frac12\right)\frac{\sqrt{\pi}\,\Gamma\left(
1-\frac{1}{R^2}\right)}{\Gamma\left(\frac32-\frac{1}{R^2}\right)}\,,
& (1<R^2<2)\, ,
\end{cases}
\label{overlineDelta}
\end{align}
where the separation is the smaller one between $\Delta$ and
$N-\Delta$.
We are considering only the region $R^2<2$ since we have
$\VEV{p}\sim N^{2-2/R^2}\gg N$ in the other region $R^2>2$.
Note that the critical $R^2$ below which the $y$-integration in the
numerator of \eqref{overlineDelta} diverges at the edges has changed
to $R^2=1$ due to the presence of $\min(y,\pi-y)$.

As seen from \eqref{overlineDelta}, the average separation
$\overline{\Delta}$ diverges as $R^2\uparrow 1$.
This is consistent with the boundstate-dissociated-state transition
which we claimed to occur at $R^2=1$.
{}From \eqref{VEVp}, \eqref{molecule} and \eqref{overlineDelta},
the condition \eqref{criterion} for the validity of the dilute pair
approximation \eqref{Z_pairs} is now explicitly given by
\be
\frac{1}{1/R^2 -1}\left(\frac{\pi e^{b/4}}{2}\right)^{-2/R^2}
\lesssim 1 \, ,
\quad (R^2<1)\, .
\label{criterionexplicit}
\ee
The breakdown temperature of the dilute pair approximation determined
by \eqref{criterionexplicit} becomes lower as $b$ is decreased.
Moreover, for smaller $b$ we have to take into account also longer
neutral boundstates such as \eqref{chain}, and hence the dilute pair
approximation becomes worse than the above estimate.
However, we would like to emphasize that the free energy argument of an
isolated charge using \eqref{Fisolatedcharge} holds independently of
the value of $b$.

\subsection{Monte Carlo analysis of internal energy and specific
  heat}
\label{ie_sh_MC}

In order to study whether the boundstate-dissociated-state phase
transition which we predicted in section \ref{BSandDS} really exists,
we have calculated using Monte Carlo method the internal energy $E$
and the specific heat $\CV$ of the system $t(x=0)$:
\begin{align}
&E=-\Drv{}{\beta}\frac{1}{N}\ln t(0)\, ,
\\
&\CV=\beta^2\frac{\p^2}{\p\beta^2}\frac{1}{N}\ln t(0)\,.
\end{align}
\vspace{-0.2cm}
\begin{figure}[htbp]
\begin{center}
\leavevmode
\put(202,45){$\bm{R^2}$}
\put(7,167){$\bm{E}$}
\put(434,44){$\bm{R^2}$}
\put(240,167){$\bm{\CV}$}
\epsfig{file=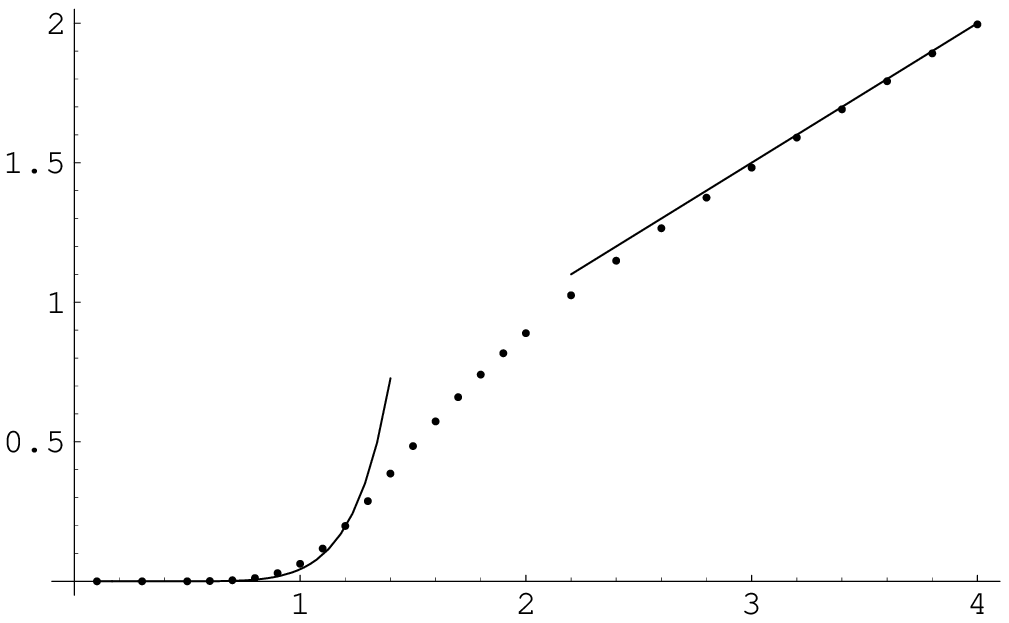, width=7cm}
\hspace*{0.8cm}
\epsfig{file=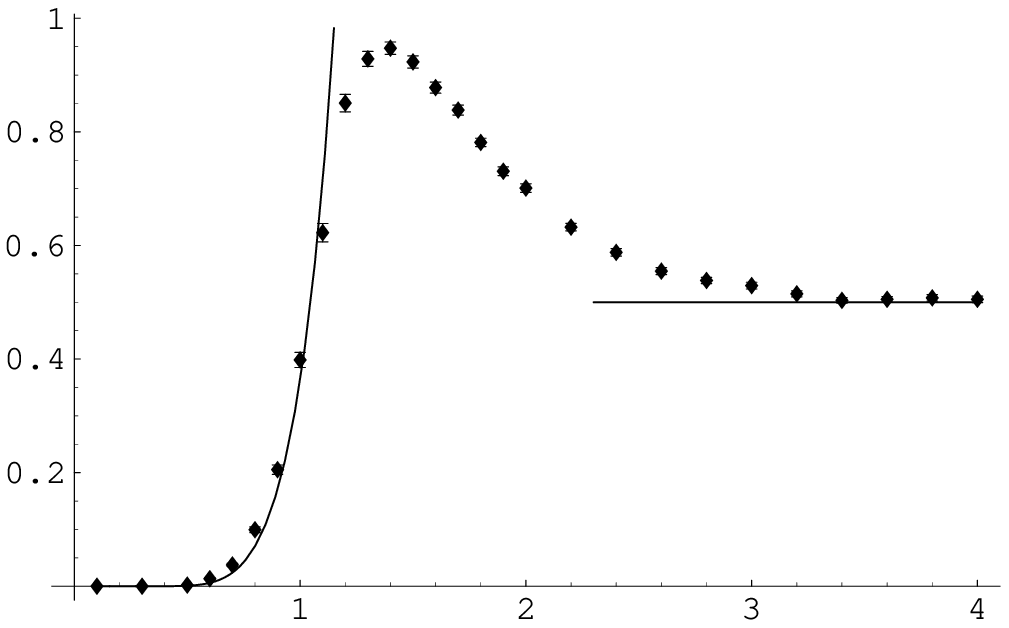, width=7cm}
\end{center}
\vspace{-1.8cm}
\caption{$E$ and $\CV$ vs.\ $R^2$ for $N=512$, $b=10$ and $x=0$.
The curves in the smaller region of $R^2$ and the straight lines in
the larger $R^2$ region have been obtained by the dilute
pair approximation and the high temperature approximation,
respectively.
}
\label{ie_cv_b10}
\end{figure}
\begin{figure}[htbp]
\begin{center}
\leavevmode
\put(202,44){$\bm{R^2}$}
\put(10,165){$\bm{E}$}
\put(436,44){$\bm{R^2}$}
\put(245,167){$\bm{\CV}$}
\epsfig{file=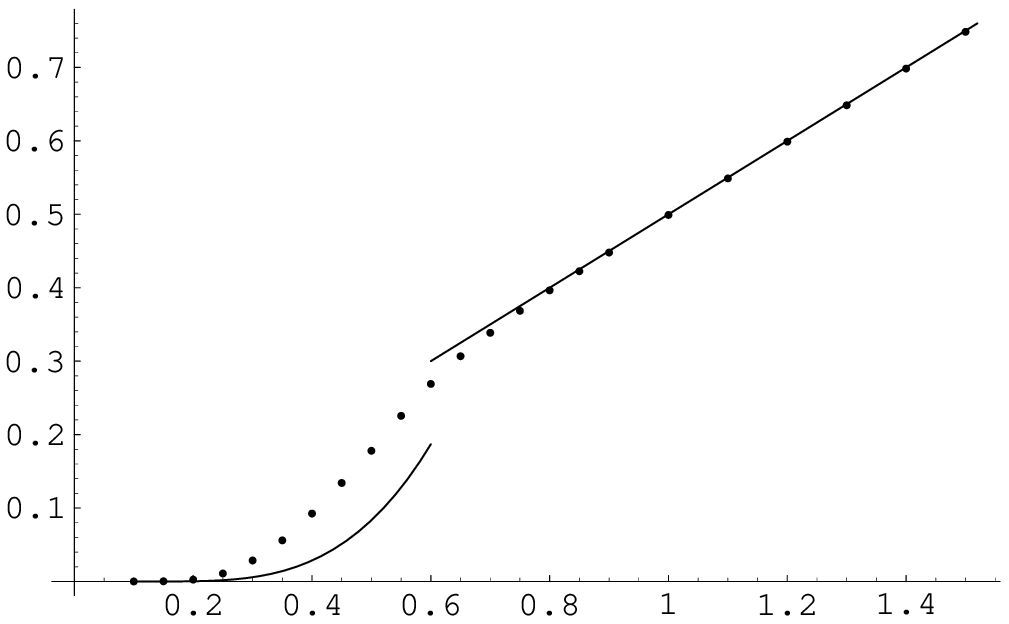, width=7cm}
\hspace*{10mm}
\epsfig{file=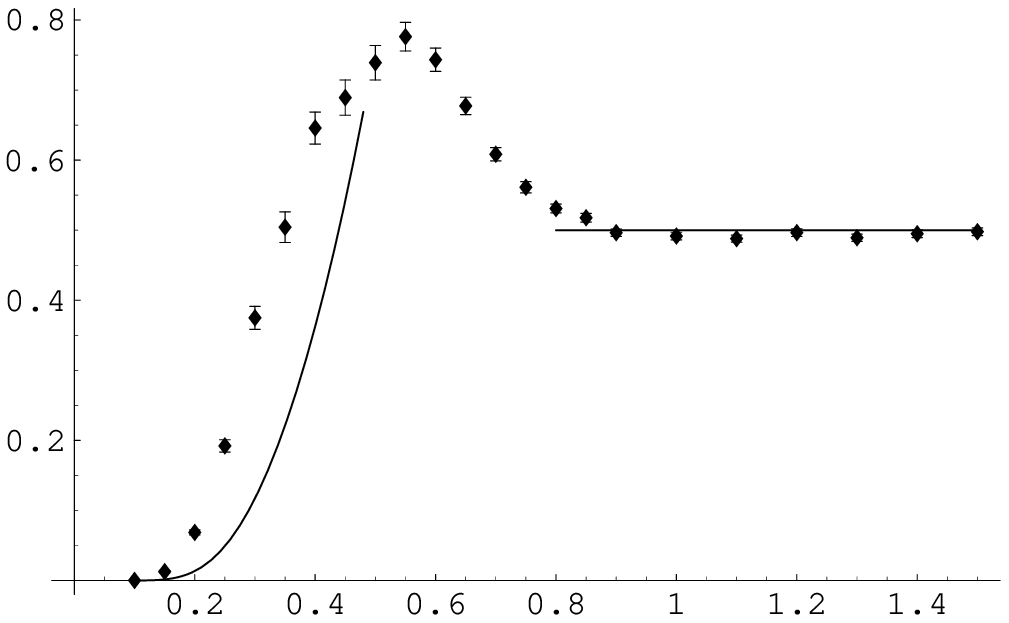, width=7cm}
\vspace{-1.5cm}
\end{center}
\caption{$E$ and $\CV$ vs.\ $R^2$ for $N=512$, $b=1$, and $x=0$.
}
\label{ie_cv_b1}
\end{figure}
Figures \ref{ie_cv_b10} and \ref{ie_cv_b1} show $E$ and $\CV$ for
$b=10$ and $b=1$, respectively ($N=512$ in both the figures).
The curves in the smaller $R^2$ region have been obtained by using the
dilute pair approximation \eqref{Z_pairs}:
\be
\frac{1}{N}\ln t(0)\simeq \frac{2}{2\beta-1}
\left(\frac{\pi e^{b/4}}{2}\right)^{-2\beta}\,,
\quad (R^2\ll 1)\, .
\ee
The staright lines in the larger $R^2$ region are from the high
temperature approximation (c.f.\ \eqref{tn0R^2gg1}):
\be
\frac{1}{N}\ln t(0)\simeq -\frac12\ln\beta\, ,
\quad (R^2\gg 1)\, ,
\ee
giving
\be
E\simeq \frac12 R^2,
\quad
\CV\simeq \frac12,\quad (R^2\gg1)\, .
\label{ECVhighR}
\ee
The curves of the dilute pair approximation fit better with the data
in the $b=10$ case than in the $b=1$ one.
This is consistent with our analysis in section \ref{secDPA} using
\eqref{criterionexplicit}.

Figures  \ref{ie_cv_b10} and \ref{ie_cv_b1} show no sign of phase
transition around $R^2=1$.
Note that there is a peak strucrure in $\CV$.
The $R^2$ of the peak becomes larger as the parameter $b$ is
increased.
We have carried out simulations for smaller values of $N$,
and found that $\CV$, and, in particular, the height of the peak are
almost indepdendent of $N$.
Therefore, the peak in $\CV$ is not a signal of a second order phase
transition.\footnote{
The global structure of $\CV$ with a peak and the asymptotic value of
$1/2$ can roughly be reproduced from a simple partititon function
$\sum_{n=-\infty}^\infty\exp\left(-\beta b n^2/4\right)$ neglecting
the Coulomb interactions although the position and the height of the
peak differ from those obtained by simulations.
}

\begin{figure}[htbp]
\begin{center}
\leavevmode
\put(202,45){$\bm{\beta}$}
\put(7,167){$\bm{E}$}
\put(432,45){$\bm{\beta}$}
\put(240,167){$\bm{\CV}$}
\epsfig{file=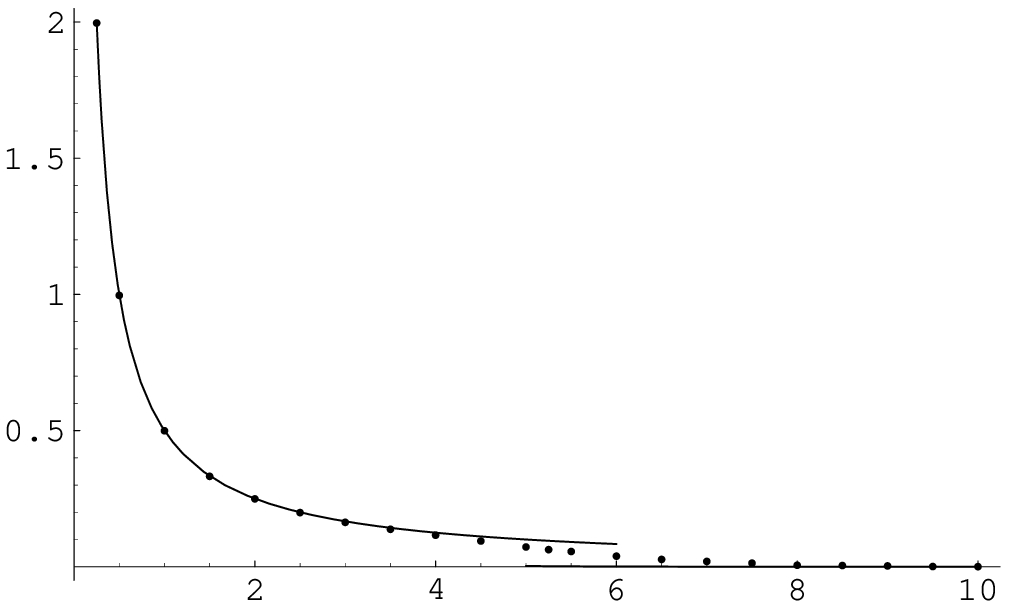, width=7cm}
\hspace*{0.8cm}
\epsfig{file=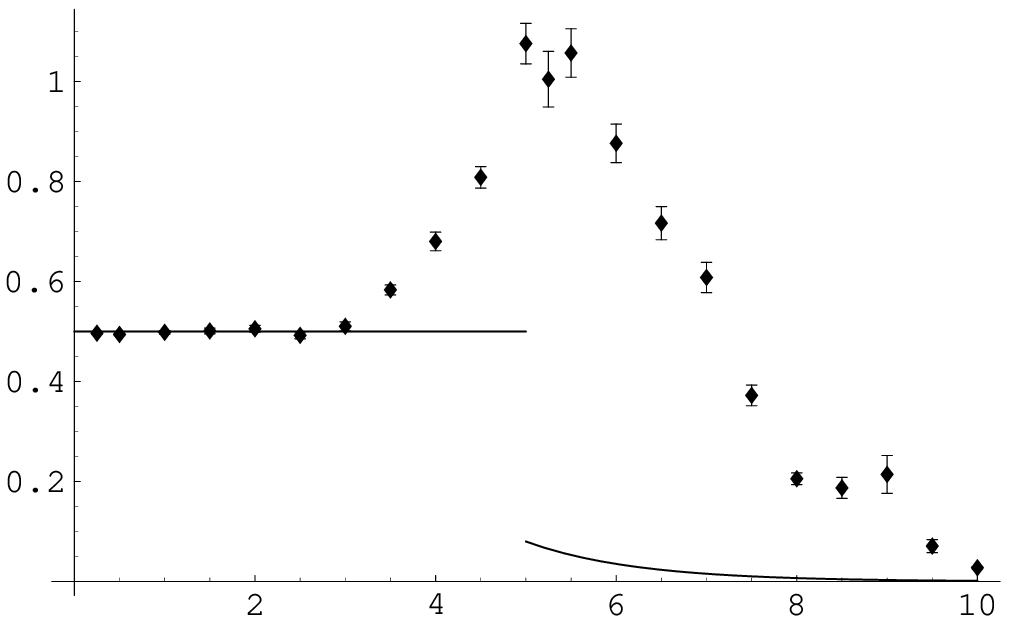, width=7cm}
\end{center}
\vspace{-1.5cm}
\caption{$E$ and $\CV$ for $b=0$ and $N=511$. Here, the horizontal
axis is $\beta$. The curve of $E$ in the dilute pair approximation in
the larger $\beta$ region is invisible since it almost overlaps with
the $\beta$-axis.
}
\label{fig:ie_cv_b0}
\end{figure}
We claimed in section \ref{solution_b0} that we have to set $b=0$ and
$N=\mbox{odd}$ for obtaining a sensible rolling solution in our
construction.
Figure \ref{fig:ie_cv_b0} shows $E$ and $\CV$ of $t(0)$ for $b=0$ and
$N=511$.
The high temperature approximation \eqref{ECVhighR} fits well with the
data in the region $\beta\lesssim 3$. However, the dilute pair
approximation is not a good approximation even
in the region of the largest $\beta$ in the figure.
As we mentioned in section \ref{secDPA}, we have to take into account
longer neutral boundstates besides the simple pair for obtaining a
better low temperature approximation.\footnote{
In fact, incorporation of the chain excitation \eqref{chain} with
$q=1$, whose energy is given by \eqref{Hchain} in the next subsection
for sufficiently large length $\Delta$, seems to considerably improve
the low temperature approximation for $b=0$.
}
In any case, we cannot observe any sign of lower order phase
transitions from the figure.

\subsection{Correlation function}

Next we shall investigate the correlation functions of $n_r$
in the low and the high tempareture regions.
Here, we consider the two-point correlation function
$\VEV{n_r\,n_{r+\Delta}}$ in the system with partition function
$t(x=0)$ for a large distance $\abs{\Delta}\gg 1\pmod N$:
\be
\VEV{n_k n_{k+\Delta}}=\frac{1}{t(0)}
\sum_{\begin{smallmatrix}
n_0,\cdots,n_{N-1}=-\infty\\
(n_0+\cdots+n_{N-1}=0)
\end{smallmatrix}}^\infty
n_k n_{k+\Delta} e^{-\beta H(n_r;n_0)}\,.
\label{correlation}
\ee

Let us consider the low temperature region $R^2\ll 1$.
There are two candidate configurations with lower energy which mainly
contribute to  \eqref{correlation}.
One is the configuration of a pair excitation, $\{\pm
n_r^{(k,\Delta)}\}$, which we defined in section \ref{BSandDS};
$(n_k,n_{k+\Delta})=(\pm 1,\mp 1)$ and all the other $n_r=0$.
The other configuration is the chain of alternating sign charges
\eqref{chain} with ends at $r=k$ and $k+\Delta$;
$(n_k,n_{k+1},\cdots,n_{k+\Delta})=\pm(1,-1,\cdots,-1,1)$ and all
other $n_r=0$. We call it the chain excitation.
This chain excitation exists only in the case of odd $\Delta$.
In the case of even $\Delta$, there are similar alternating sign
configurations with zero total charge. Here, for simplicity, we
consider only the case of odd $\Delta$.

\begin{figure}[htbp]
\vspace{-1.0cm}
\begin{center}
\leavevmode
\put(12,162){$\bm{H_{\rm chain}}$}
\put(205,53){$\bm{\Delta}$}
\epsfig{file=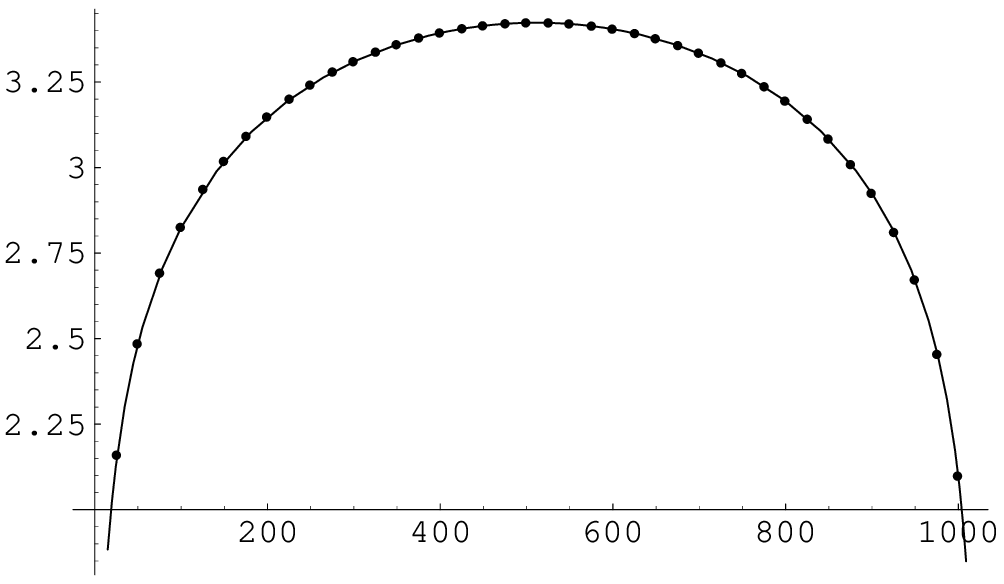, width=7cm}
\vspace{-1.8cm}
\end{center}
\caption{The energy of the chain excitation at various $\Delta$
in the case of $N=1023$ and $b=0$ (dots).
The curve is $H_{\rm chain}=3.42 + 0.500 \ln[\sin(\pi\Delta/N)]$
obtained by fitting.
}
\label{e_pm1}
\end{figure}

The energy of the pair excitation is given by \eqref{energy}.
We calculated numerically the energy  of the chain excitation
$H_{\rm chain}$. The dots in figure \ref{e_pm1} show $H_{\rm chain}$
for the various lengths $\Delta$ ($N=1023$ and $b=0$).
These points are well fitted by the curve
$H_{\rm chain}=3.42 + 0.500 \ln[\sin(\pi\Delta/N)]$.
We have carried out this analysis for various values of $N$, and found
that the dependence of $H_{\rm chain}$ on (sufficiently large)
$\Delta$ and $N$ is given by
\be
H_{\rm chain}\bigr|_{b=0}=\frac12\ln N +
\frac12\ln\!\left[\sin\left(\frac{\pi\Delta}{N}\right)\right]
+\mbox{const.}\,.
\label{Hchain}
\ee
In particular, the coefficient of $\ln\sin(\pi\Delta/N)$ is $1/2$ and
independent of $N$.
In the case of $b\neq 0$, $H_{\rm chain}$ has an additional
self-energy contribution $(b/4)(\Delta+1)$.
Thus we find that the contributions of the two kinds of excitations
to the correlation function are
\begin{numcases}
{\VEV{n_k\,n_{k+\Delta}}\sim}
\ds N^{-2/R^2}\abs{\sin\left(\frac{\pi\Delta}{N}\right)}^{-2/R^2}
&\mbox{: pair excitation}\,,
\label{nnpair}
\\
\ds N^{-1/(2R^2)}e^{-b\,\Delta/(4R^2)}
\abs{\sin\left(\frac{\pi\Delta}{N}\right)}^{-1/(2R^2)}
&\mbox{: chain excitation}\,.
\label{nnchain}
\end{numcases}
Eqs.\ \eqref{nnpair} and \eqref{nnchain} are valid in a wide range of
$\Delta$ including both $\Delta=O(N/2)$ and $1\ll \Delta\ll N/2$.
In particular, in the region $1\ll \Delta\ll N/2$, the $N$-dependence
cancels out and we obtain
\begin{numcases}
{\VEV{n_k\,n_{k+\Delta}}\underset{1\ll \Delta\ll N/2}{\sim}}
\ds \frac{1}{\Delta^{2/R^2}}
&\mbox{: pair excitation}\,,
\label{nnpairmedD}
\\
\ds e^{-b\,\Delta/(4R^2)}\frac{1}{\Delta^{1/(2R^2)}}
&\mbox{: chain excitation}\,.
\label{nnchainmedD}
\end{numcases}
{}From \eqref{nnpair}--\eqref{nnchainmedD} we see the
followings. In the case of $b\ne 0$, $\VEV{n_k\,n_{k+\Delta}}$
is given by the contribution from the pair excitation, \eqref{nnpair}
and \eqref{nnpairmedD}, for a large distance $\Delta\gg
1$. Contribution of the chain excitation is suppressed by
$e^{-b\,\Delta/(4R^2)}$.
On the other hand, in the case of $b=0$, $\VEV{n_k\,n_{k+\Delta}}$
is given by the contribution from the chain excitation,
\eqref{nnchain} and \eqref{nnchainmedD}.

Next, let us consider the high temperature region $R^2\gg 1$.
For $R^2\gg 1$, approximating the summations over $n_r$ in
\eqref{correlation} by integrations over continuous variables
$p_r=n_r/R$, we obtain
\be
\VEV{n_k\,n_{k+\Delta}}
\sim R^2 \bigl(\wh{Q}^{-1}\bigr)_{k,k+\Delta}\,.
\ee
Incidentally, $\bigl(\wh{Q}^{-1}\bigr)_{k,k+\Delta}$ with $k=0$ is
related to  $n^C_r$ \eqref{nrC} by\footnote{
Numerical analysis shows that
$\det\hhQ/\det\wh{Q}$ is finite in the limit $N\to\infty$ for $b\ne0$,
while we have $\det\hhQ/\det\wh{Q}\simeq 0.59\,N$ for $b=0$.
}
\be
\bigl(\wh{Q}^{-1}\bigr)_{0\Delta}
=\frac{\det\hhQ}{\det\wh{Q}}\,n^C_\Delta \, .
\ee
We calculated $(\wh{Q}^{-1})_{k,k+\Delta}$ numerically and found
that it is independent of the position $k$; namely, the translational
invariance holds for large $N$ despite that the self-energy $b/4$ is
missing for $n_0$ (see \eqref{Qrs}).
Therefore, $\bigl(\wh{Q}^{-1}\bigr)_{k,k+\Delta}$ is essentially equal
to $n^C_\Delta$ up to a $\Delta$-independent factor.

The $\Delta$-dependence of $n_\Delta^C$ is shown in figure
\ref{fig:n_r^C} in the cases of $b=0.1$ and $10$.
As we mentioned in the footnote there, $n_\Delta^C$ and hence
$(\wh{Q}^{-1})_{k,k+\Delta}$ has quite different $\Delta$-dependences
in the smaller $\Delta$ region: for larger $b$,
$(\wh{Q}^{-1})_{k,k+\Delta}$ is negative definite, while it has
alternating sign structure for smaller $b$.
However, for $\Delta=O(N/2)$ in the mid region,
$(\wh{Q}^{-1})_{k,k+\Delta}$ is negative definite and has the
following universal $\Delta$-dependence for any non-vanishing $b$ as we
shall see below:
\be
(\wh{Q}^{-1})_{k,k+\Delta}\sim
N^{-2}\abs{\sin\left(\frac{\pi\Delta}{N}\right)}^{-2}\,,
\quad (\Delta=O(N/2))\, .
\label{commonQinv}
\ee

\begin{figure}[htbp]
\vspace{-0.8cm}
\begin{center}
\leavevmode
\put(203,82){$\bm{\Delta}$}
\put(5,165){$\bm{\ln|(\wh{Q}^{-1})_{0\Delta}|}$}
\epsfig{file=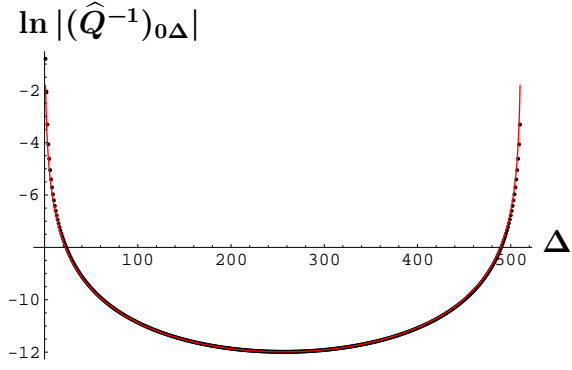, width=7.0cm}
\vspace{-1.8cm}
\end{center}
\caption{$\ln|(\wh{Q}^{-1})_{0\Delta}|$ for $N=511$ and $b=10$ (dots)
and the fitted red curve $-12.02-2.107\ln\abs{\sin(\pi\Delta/N)}$.
}
\label{fig:qhatinvb10}
\end{figure}
\begin{figure}[htbp]
\begin{center}
\leavevmode
\put(200,95){$\bm{\Delta}$}
\put(5,168){$\bm{(\wh{Q}^{-1})_{0\Delta}}$}
\put(431,95){$\bm{\Delta}$}
\put(228,165){$\bm{\ln|(\wh{Q}^{-1})_{0\Delta}|}$}
\epsfig{file=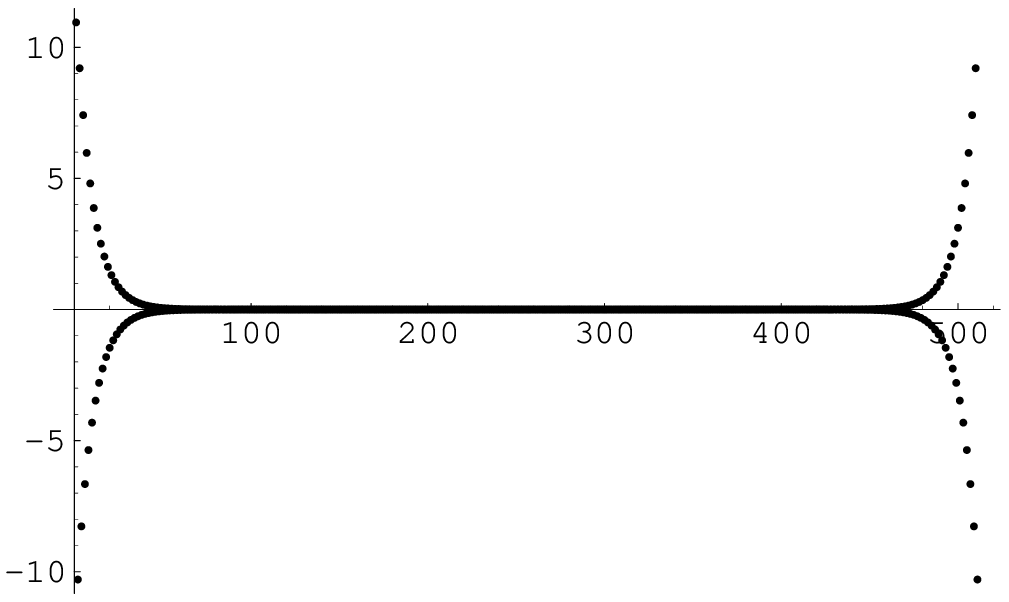, width=7cm}
\hspace*{0.8cm}
\epsfig{file=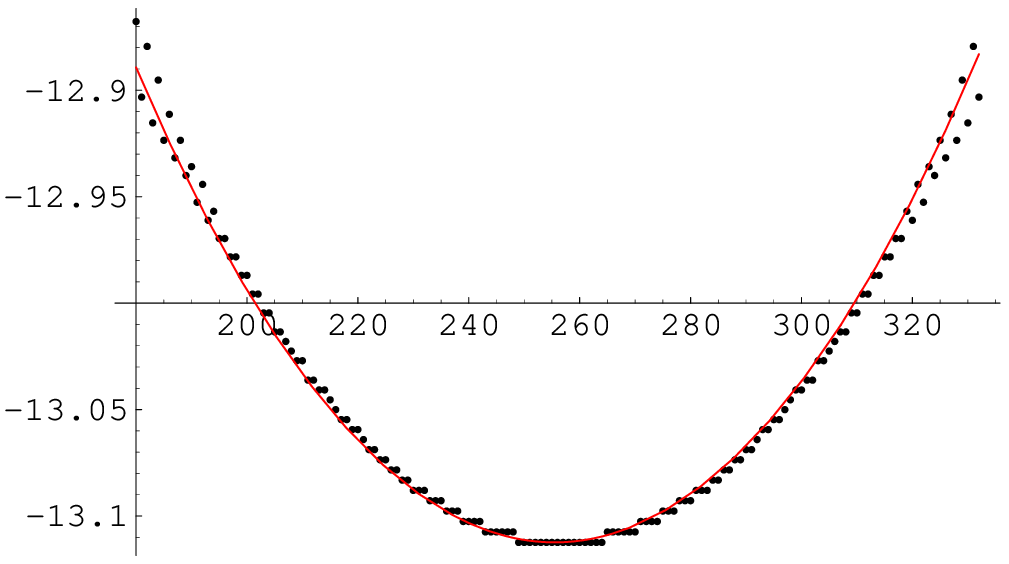, width=7cm}
\end{center}
\vspace{-1.8cm}
\caption{
The left figure shows $\bigl(\wh{Q}^{-1}\bigr)_{0\Delta}$ for
$b=0.01$ and $N=511$ in all the region $\Delta=1,\cdots,N-1$.
The right figure shows $\ln|\bigl(\wh{Q}^{-1}\bigr)_{0\Delta}|$
and the fitted red curve
$-13.09-2.00\ln\abs{\sin(\pi\Delta/N)}$
only in the mid region $180\le\Delta\le 332$.
}
\label{qhatinv_511_0.01}
\end{figure}

First, for larger $b$, $(\wh{Q}^{-1})_{k,k+\Delta}$ is well fitted
by \eqref{commonQinv} for any $\Delta$ (see figure
\ref{fig:qhatinvb10}). We have confirmed the $N$-dependence of
\eqref{commonQinv} by the fitting for various $N$.
Next, let us consider $(\wh{Q}^{-1})_{k,k+\Delta}$ for a smaller $b$.
Figure \ref{qhatinv_511_0.01} shows $(\wh{Q}^{-1})_{0\Delta}$ in the
case of $b=0.01$ and $N=511$ in all the region of $\Delta$ (left
figure), and $\ln|(\wh{Q}^{-1})_{0\Delta}|$ in the mid region
$180\le\Delta\le 332$ where $(\wh{Q}^{-1}\bigr)_{0\Delta}$ is negative
definite (right figure).
As we see from the right figure, the $\Delta$-dependence of
\eqref{commonQinv} holds well in the mid region of $\Delta$.
By carrying out this analysis for various $N$, we have confirmed
the $N$-dependence of \eqref{commonQinv} also for smaller $b$.

\begin{figure}[htbp]
\begin{center}
\leavevmode
\put(217,105){$\bm{\Delta}$}
\put(5,180){$\bm{(\wh{Q}^{-1})_{0\Delta}}$}
\epsfig{file=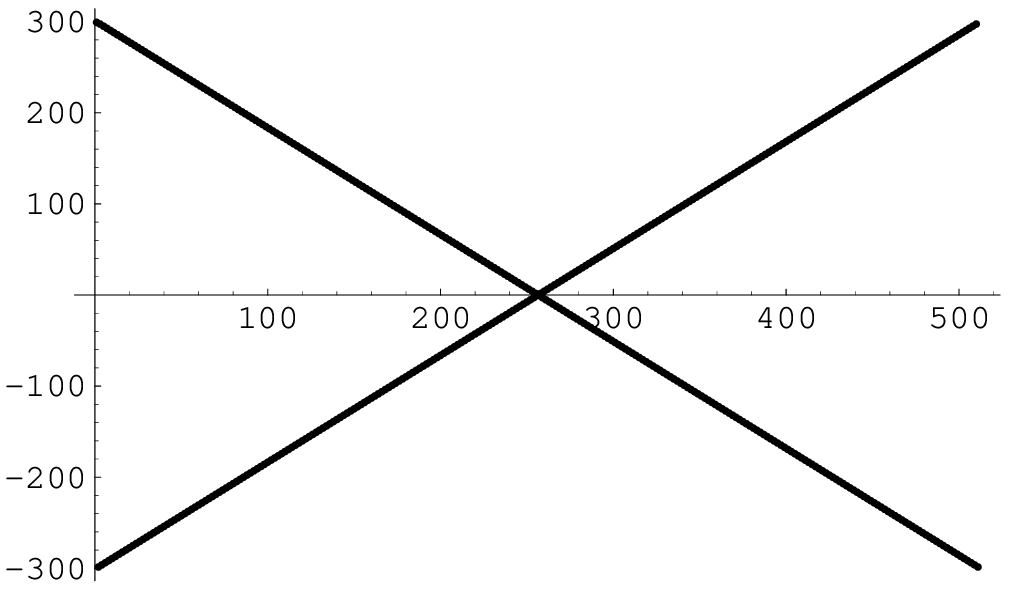, width=7.5cm}
\vspace{-1.8cm}
\end{center}
\caption{$\bigl(\wh{Q}^{-1}\bigr)_{0\Delta}$ for $N=511$ and $b=0$.}
\label{qhatinv_511_0}
\end{figure}

The behavior of $\bigl(\wh{Q}^{-1}\bigr)_{0\Delta}$ in the case of
$b=0$ and $N=\mbox{odd}$ is quite different from the non-zero $b$ case
above. $n^C_r$ in this case is given by \eqref{zero_odd}, and
$\bigl(\wh{Q}^{-1}\bigr)_{0\Delta}$ is shown in figure
\ref{qhatinv_511_0} for $N=511$ and $b=0$.
$(\wh{Q}^{-1})_{0\Delta}$ is a linear function of $\Delta$ with
alternating sign for all $\Delta$.

{}From the above analysis of $\VEV{n_k\,n_{k+\Delta}}$ in both the
$R^2\ll 1$ and $R^2\gg 1$ regions, we find the followings.
In the case of $b\ne 0$, $\VEV{n_k\,n_{k+\Delta}}$ is given by
\eqref{nnpair} in the $R^2\ll 1$ region, and by \eqref{commonQinv} in
the $R^2\gg 1$ region. Namely,
$\VEV{n_k\,n_{k+\Delta}}$ has the same kind of $\Delta$-dependence
$|\sin(\pi\Delta/N)|^{-\gamma}$ although the exponent $\gamma$ differs
in the two regions. On the other hand, $\VEV{n_k\,n_{k+\Delta}}$ in
the $b=0$ case has entirely different $\Delta$-dependences in the two
regions; it is given by \eqref{nnchain} in the $R^2\ll 1$ region, and
by \eqref{zero_odd} in the $R^2\gg 1$ region. This result suggests the
existence of a phase transition at an intermediate $R^2$ at least in
the $b=0$ case.
Further study of the correlation function in the intermediate region
of $R^2$ using Monte Carlo simulation is needed.

\section{Summary and Discussions}

In this paper, we constructed a time-dependent solution in VSFT
and studied whether it can represent the rolling tachyon process.
Our solution is given as the inverse Wick rotation of the lump
solution on a circle of raduis $R$ which is given as an infinite
number of $*$-products $\Omega_b *\Omega_b *\cdots *\Omega_b$
of a string field $\Omega_b$ with Gaussian momentum dependence
$e^{-b\,p^2/4}$.
We focused on one particular component field in the solution, which
has an interpretation as the partition function of a Coulomb system on
a circle with temperature $R^2$.
Our finding in this paper is that, for the solution not to diverge in
the large time limit, we have to put $b=0$ and take the number of
$\Omega_b$ consituting the solution to infinity by keeping it even.
We also examined the various thermodynamic quantities of our solution
as a Coulomb system to see whether the self-dual radius $R=1$ has a
special meaning.
We pointed out a possibility that $R=1$ is a phase transition point
separating the boundstate phase and the dissociated state phase.
Our analysis of the correlation function for $b=0$ supports the
existence of the phase transition.

Many parts of this paper are still premature and need further study.
The most important among them is to study in more detail our solution
with $b=0$: whether the limit $N\to\infty$ really exists, and if so,
what the profile will be. In this paper we tried to find a special
nature of the critical radius $R=1$ in the thermodynamic properties of
the system. However, we have to find a more direct relevance of $R=1$ to
our solution. For example, the most natural scenario is that the limit
$N\to\infty$ can exist only at $R=1$. Analysis of the whole of our
solution not restricted to the component $t(x^0)$ is also necessary.

Originally our time-dependent solution had two parameters, $b$ and
$R$. If we have to put $b=0$ and $R=1$ for obtaining a solution with a
desirable rolling profile, this solution seems to have no free
parameters. However, the rolling solution should have one free
parameter which corresponds to the initial tachyon value at $x^0=0$.
It is our another problem to find the origin of this parameter.
It might be necessary to generalize our solution to incorporate this
parameter.

After establising the rolling solution, our next task is of course to
apply our solution to the analysis of unresolved problems in the
rolling tachyon physics.

\section*{Acknowledgments}

We would like to thank H.\ Fukaya, M.\ Fukuma, Y.\ Kono, T.\ Matsuo,
K.\ Ohmori, S.\ Shinomoto, S.\ Teraguchi and E.\ Watanabe for valuable
discussions. The work of H.\,H. was supported in part by a
Grant-in-Aid for Scientific Research from Ministry of Education,
Culture, Sports, Science, and Technology (\#12640264).

\begin{appendix}

\section{Proof of eq.\ (\ref{whQnC=1/N})}

In this appendix we present a proof of \eqref{whQnC=1/N} for
$n_r^C$ given by the RHS of \eqref{zero_odd}.
Since this $n_r^C$ satisfies $\sum_{r=0}^{N-1}n_r^C=0$,
and for such $n_r^C$ we have
\begin{equation}
\sum_{s=0}^{N-2}\wh{Q}_{rs}n_r^C
=\sum_{s=0}^{N-1}Q_{rs}n_s^C-\sum_{s=0}^{N-1}Q_{N-1,s}n_s^C \, ,
\end{equation}
eq.\ \eqref{whQnC=1/N} holds if we can show that
\begin{equation}
\sum_{s=0}^{N-1}Q_{rs}n_s^C=(\mbox{$r$-independent term})
+O\!\left(\frac{1}{N}\right)\, ,
\quad (r=0,1,\cdots,N-1) \, .
\label{eq:goal}
\end{equation}

Before starting the proof of \eqref{eq:goal} for $n_r^C$ of
\eqref{zero_odd}, we shall mention eq.\ \eqref{hatQ(-1)^s=0} in the
$b=0$ and $N=\mbox{even}$ case.
This \eqref{hatQ(-1)^s=0} holds owing to a stronger equation
$\sum_{s=0}^{N-1}Q_{rs}(-1)^s=0$, which is rewritten explicitly as
\begin{equation}
\sum_{r=1}^{N-1}(-1)^r\ln\abs{2\sin\!\left(\frac{\pi r}{N}\right)}
=\ln\!\left(\frac{N}{4}\right)\, ,
\quad (N=\mbox{even})\, .
\label{eq:goaleven}
\end{equation}
Formulas essentially equivalent to \eqref{eq:goaleven} can be found in
the various tables of series and products.

Now let us consider the LHS of \eqref{eq:goal} for $n_r^C$ given by
the RHS of \eqref{zero_odd}.
Taylor expanding
$\ln\abs{2\sin\left[\pi(r-s)/N\right]}
=\Re\ln\!\left(1-e^{2\pi i(r-s)/N}\right)$ in $Q_{rs}$ \eqref{Qrs}
in power series of $e^{2\pi i(r-s)/N}$, we have
\begin{align}
\frac{(-1)^r}{2}\!\!\!
\sum_{s=-\frac{N-1}{2}}^{\frac{N-1}{2}}\!\!Q_{rs}n^C_s
&=\ln\!\left(\frac{N}{4}\right)
\left(1-\frac{2\abs{r}}{N+1}\right)
+\Re\sum_{k=1}^\infty\frac{1}{k}\!\!
\sum_{\begin{smallmatrix}s=-\frac{N-1}{2}\\(\ne r)
\end{smallmatrix}
}^{\frac{N-1}{2}}\!\!\left(-e^{2\pi i k/N}\right)^{r-s}
\left(1-\frac{2\abs{s}}{N+1}\right)
\nn\\
&
=\ln\!\left(\frac{N}{4}\right)
\left(1-\frac{2\abs{r}}{N+1}\right)
+\sum_{k=1}^\infty\frac{1}{k}\left[
f_k -\left(1-\frac{2\abs{r}}{N+1}\right)\right]\, ,
\label{eq:stepone}
\end{align}
with $f_k$ defined by
\begin{equation}
f_k=-\frac{(-1)^r}{N+1}\cdot\frac{
(-1)^{\frac{N+1}{2}+k}\left[
e^{i\pi k/N} + e^{-i\pi k/N}\right]-2}{
\left(e^{i\pi k/N}+e^{-i\pi k/N}\right)^2}
\times\left[\left(e^{2\pi i k/N}\right)^r
+ \left(e^{-2\pi i kr/N}\right)^r\right]\, .
\label{eq:f_k}
\end{equation}
In obtaining the last line of \eqref{eq:stepone} we have
applied the formula
\begin{equation}
\sum_{s=-M+1}^{M-1}z^{-s}\left(1-\frac{\abs{s}}{M}\right)
=\frac{1}{M}\cdot\frac{z(z^M +z^{-M}-2)}{(1-z)^2}\, ,
\end{equation}
to the case of $M=(N+1)/2$ and $z=-e^{2\pi i k/N}$, and
used that
$\left(-e^{\pm 2\pi i k/N}\right)^{\frac{N+1}{2}}
=(-1)^{\frac{N+1}{2}}\times\left(-e^{\pm i \pi/N}\right)^k$.

Note that $f_k$ defined above has the periodicity:
\begin{equation}
f_{k+N}=f_k \, .
\label{eq:propf_k}
\end{equation}
Introducing the cutoff $LN$ in the $k$-summation in
\eqref{eq:stepone} and using the periodicity to make the manipulation
$\sum_{k=1}^{LN}f_k/k=\sum_{k=1}^N f_k\sum_{p=0}^{L-1}1/(pN+k)$,
we have
\begin{align}
&\sum_{k=1}^{LN}\frac{1}{k}\!\left[f_k
-\left(1-\frac{2\abs{r}}{N+1}\right)\right]
\nn\\
&=\frac{1}{N}\sum_{k=1}^N f_k\left[
\psi\!\left(L+\frac{k}{N}\right)-\psi\!\left(\frac{k}{N}\right)
\right]
-\left(1-\frac{2\abs{r}}{N+1}\right)\bigl[\psi(LN+1)-\psi(1)\bigr]
\nn\\
&\underset{L\gg 1}{=}
\frac{1}{N}\sum_{k=1}^N f_k\left[
\ln L-\psi\!\left(\frac{k}{N}\right)
\right]
-\left(1-\frac{2\abs{r}}{N+1}\right)\left[\ln(LN)+\gamma\right]\, ,
\label{eq:steptwo}
\end{align}
where $\psi(z)$ is the polygamma function:
\begin{equation}
\psi(z)=\Drv{}{z}\ln\Gamma(z)
=-\gamma-\sum_{n=0}^\infty\left(\frac{1}{n+z}-\frac{1}{n+1}\right)\,.
\end{equation}
In obtaining the last line of \eqref{eq:steptwo}, we have used
$\psi(1)=-\gamma$ and the asymptotic behavior of $\psi(z)$ for
$\abs{z}\gg 1$:
\begin{align}
\psi(z)&\simeq\ln(z-1)+\frac{1}{2(z-1)}-\frac{1}{12(z-1)^2}
+\ldots \, .
\end{align}

Now we have to carry out the two summations in \eqref{eq:steptwo},
$$
S_1=\frac{1}{N}\sum_{k=1}^N f_k\, ,
\qquad
S_2=\frac{1}{N}\sum_{k=1}^N f_k\,\psi\!\left(\frac{k}{N}\right)
\, ,
$$
for large $N$.
One way to evaluate $S_1$ is to approximate it by a contour
integration with respect to $z=e^{2\pi i k/N}$:
\begin{equation}
S_1\underset{N\gg 1}{=}
\frac{1}{2\pi i}\oint\limits_{\abs{z}=1}\frac{dz}{z}\,f(z)
=\Res\limits_{z=0}\frac{1}{z}f(z)
=1-\frac{2\abs{r}}{N+1}\, ,
\label{eq:S_1contourint}
\end{equation}
where $f(z)$ is
\begin{equation}
f(z)=\frac{-z}{N+1}\cdot
\frac{(-z)^{\frac{N+1}{2}}+(-1/z)^{\frac{N+1}{2}}-2}{
(1+z)^2}\cdot\Bigl[(-z)^r +(-1/z)^r\Bigr]\, .
\end{equation}
Note that $f(z)/z$ is regular at $z=-1$.

Another way of evaluating $S_1$ is to observe the followings.
$f_k$ is of $\calO(1/N)$ except at $k\sim (N+1)/2$ where $f_k=\calO(N)$.
Therefore, we have only to carry out the $k$-summation around
$k\sim (N+1)/2$.
Expressing $k$ as $k=(N+1)/2+\ell$, $f_k$ is expanded around
$k=(N+1)/2$ as
\begin{equation}
f_{k=(N+1)/2+\ell}=\frac{1}{N+1}\left\{
\frac{N^2}{\pi^2\left(\ell+\frac12\right)^2}
+\frac{N(-1)^\ell}{\pi\left(\ell+\frac12\right)}
+O(1)\right\}
\cos\!\left[\frac{2\pi r}{N}\!\left(\ell+\frac12\right)\right]\, ,
\label{eq:expandf_k}
\end{equation}
and we regain the same result as \eqref{eq:S_1contourint}:
\begin{equation}
S_1\simeq\frac{1}{N}\sum_{\ell=-\infty}^\infty\eqref{eq:expandf_k}
=\frac{1}{N+1}\left\{
N\left(1-\frac{2\abs{r}}{N}\right)+1\right\}
=1-\frac{2\abs{r}}{N+1} \, ,
\label{eq:evalS_1}
\end{equation}
where we have used the formulas,
\begin{align}
&\sum_{n=0}^\infty\frac{\cos(2n+1)x}{(2n+1)^2}
=\frac{\pi}{4}\left(\frac{\pi}{2}-\abs{x}\right)\, ,
\\
&\sum_{n=0}^\infty (-1)^{n}\frac{\cos(2n+1)x}{2n+1}
=\frac{\pi}{4}\, ,
\quad
\left(-\frac{\pi}{2}< x <\frac{\pi}{2}\right)\, .
\end{align}

The other summation $S_2$ is evaluated in the same manner:
\begin{equation}
S_2\simeq\sum_{\ell=-\infty}^\infty\eqref{eq:expandf_k}
\left[
\psi\!\left(\frac12\right)+\frac{1}{N}\!\left(\ell+\frac12\right)
\psi'\!\left(\frac12\right)\right]
=-\left(\gamma+\ln 4\right)\left(1-\frac{2\abs{r}}{N+1}\right)\, ,
\label{eq:evalS_2}
\end{equation}
where we have used that $\psi(1/2)=-\gamma-\ln 4$, and that
the summations
$$
\sum_{\ell=-\infty}^\infty\left\{\frac{1}{\ell+\frac12},
(-1)^\ell\right\}
\cos\!\left[\frac{2\pi r}{N}\!\left(\ell+\frac12\right)\right]\, ,
$$
vanish due to the oddness under $\ell\to -\ell-1$.
Recalling that
\begin{align}
&\frac{(-1)^r}{2}\!\!\!
\sum_{s=-\frac{N-1}{2}}^{\frac{N-1}{2}}\!\!Q_{rs}n^C_s
\nn\\
&=\ln\!\left(\frac{N}{4}\right)
\cdot\left(1-\frac{2\abs{r}}{N+1}\right)
+S_1\cdot\ln L -S_2
-\left(1-\frac{2\abs{r}}{N+1}\right)\left[\ln(LN)+\gamma\right]\, ,
\label{eq:stepthree}
\end{align}
and plugging the results \eqref{eq:evalS_1}
(or \eqref{eq:S_1contourint}) for $S_1$ and \eqref{eq:evalS_2} for
$S_2$ into \eqref{eq:stepthree}, we find that the terms on the RHS
just cancel out.
Therefore, the $r$-independent constant on the RHS of \eqref{eq:goal}
is in fact equal to zero.

\end{appendix}

\end{document}